\documentclass[conf]{new-aiaa}
\usepackage[utf8]{inputenc}
\usepackage[colorinlistoftodos]{todonotes}
\usepackage{graphicx}
\usepackage{amsmath}
\usepackage[version=4]{mhchem}
\usepackage{siunitx}
\usepackage{caption}
\usepackage{subcaption}
\usepackage{longtable,tabularx}
\usepackage{setspace}
\usepackage{cancel}
\usepackage{float}
\usepackage{multirow}
\usepackage{xcolor}
\usepackage{lineno}

\title{Flamelet Model with Epsilon Tracking in a Turbine Stator}

\author{Sylvain L. Walsh\footnote{Ph.D student, Department of Mechanical and Aerospace Engineering, AIAA Member.}, Yalu Zhu\footnote{Assistant Specialist, Department of Mechanical and Aerospace Engineering, AIAA Member.}, Feng Liu\footnote{Professor, Department of Mechanical and Aerospace Engineering, AIAA Fellow.} and William A. Sirignano\footnote{Distinguished Professor, Department of Mechanical and Aerospace Engineering, AIAA Honorary Fellow.}}
\affil{University of California, Irvine, Irvine, CA, 92697}

\begin{document}

\maketitle
 % \setstretch{1.5}
\begin{abstract}
Combustion within a two-dimensional turbine stator passage is numerically investigated in the context of the turbine-burner concept using a Reynolds-Averaged Navier-Stokes framework coupled with a novel flamelet model. The formulation links resolved-scale turbulence quantities with subgrid flamelet dynamics through the local turbulent kinetic energy dissipation rate, $\boldsymbol{\epsilon}$, which determines the flamelet inflow strain rate. For the first time, combustion of JP-5 is considered in a turbine stator passage as a practical fuel. This is achieved by solving transport equations for 14 major species on the resolved scale, while chemical source terms are obtained from precomputed flamelet libraries based on the HyChem A3 mechanism comprising 119 species and 841 elementary reactions. Model performance is assessed against methane combustion using both a one-step kinetics model and an $\boldsymbol{\epsilon}$-based flamelet formulation employing a 13-species skeletal mechanism. The $\boldsymbol{\epsilon}$-based formulation predicts lower peak flame temperatures due to dissociation effects and approximately 50\% lower net chemical energy addition per unit mass compared with the one-step model, as a result of flame stand-off and downstream strain-rate-induced quenching. For JP-5, the simulations capture combined endothermic pyrolysis and exothermic oxidation processes, leading to vertically displaced reaction zones, increased near-wall temperatures, and larger resolved-scale reaction regions due to the higher flamelet flammability limit relative to methane.

\end{abstract}

\section*{Nomenclature}
{\renewcommand\arraystretch{1.0}
\noindent\begin{longtable*}{@{}l @{\quad=\quad} l@{}}
$C_{\mu}$ & $k-\epsilon$ turbulence model empirical constant\\
$C_{\chi}$ &  proportionality constant between turbulence and scalar time-scales\\
$C_p$ & specific heat capacity at constant pressure \\
$E$ & total sensible energy \\
$h$ & sensible enthalpy\\
$h^0_f$ & enthalpy of formation \\
$\mathbf{I}$ & Kronecker tensor \\
$\mathbf{j}$ & diffusive flux \\
$k$ & turbulent kinetic energy \\
$N$ & total number of species\\
$p$ & pressure \\
$\mathrm{Pr}$ & Prandtl number\\
$\dot{Q}$ & heat release rate\\
$R$ & gas constant \\
$R_0$ & universal gas constant, $R_0$ = 8.3145 J/mol/K \\
$\mathrm{Sc}$ & Schmidt number\\
$T$ & temperature\\
$T_{\mathrm{ref}}$ & reference temperature, $T_{\mathrm{ref}}=298.15$ K \\
$t$ & time\\
$\mathbf{V}$ & velocity vector\\
$W$ & molecular weight \\
$Y$ & mass fraction\\
$Z$ & mixture fraction\\
$Z^{''2}$ & variance of the mixture fraction\\
$\chi$ & instantaneous scalar dissipation rate\\
$\epsilon$ & turbulent kinetic energy dissipation rate\\
$\mu$ & viscosity coefficient\\
$\omega$ & specific turbulent dissipation rate \\
$\dot{\omega}$ & production rate \\
$\psi$ & reactive scalar\\
$\rho$ & density \\
$\boldsymbol{\tau}$ & viscous stress tensor \\

\end{longtable*}}
\textit{Subscripts}
{\renewcommand\arraystretch{1.0}
\noindent\begin{longtable*}{@{}l @{\quad=\quad} l@{}}
$n$ & species specific quantity\\
$st$ & stoichiometric composition\\
$T$ & turbulent quantity\\
\end{longtable*}}

\textit{Superscripts}
{\renewcommand\arraystretch{1.0}
\noindent\begin{longtable*}{@{}l @{\quad=\quad} l@{}}
$-$ & Reynolds-averaged quantity\\
$\thicksim$ & Favre-averaged quantity\\
\end{longtable*}}
% \setstretch{2}

\section{Introduction}
\lettrine{F}{low} in a turbine is accelerating while power is extracted by the rotor. Consequently, it is possible to add heat without raising the ﬂow temperature beyond the turbine-blade material limit. Turbine-burner designs utilize this fact to increase performance by extending the combustion process from the combustion chamber into the turbine section. Thermodynamic analysis performed by Sirignano and Liu \cite{sirignano_performance_1999,liu_turbojet_2001} shows that this concept design allows for: 1) reduction in after-burner length and weight, 2) reduced specific fuel consumption, while increasing specific thrust, 3) widening the operational range of flight Mach number and compressor pressure ratio, and 4) a decrease in pollutant formation and heat transfer losses due to the reduction of peak temperatures resulting from flow acceleration. For readers unfamiliar with the turbine-burner concept, Yin and Gao \cite{yin_review_2020} conducted an extensive review of gas turbine engines with Inter-stage Turbine Burner (ITB).

The characteristics associated with augmentative combustion in a turbine-burner are attractive to jet engine designers. Nonetheless, major challenges related to aerodynamics and combustion must be addressed. Specifically, in a turbine passage, the turbulent compressive flow experiences a strong pressure gradient caused by the turbine blades, which accelerates the flow from subsonic to supersonic within a relatively short distance. These conditions result in millisecond residence times and flow accelerations on the order of $\sim10^5$ g, both of which pose difficulties for flameholding \cite{sirignano_turbine_2012}. Additionally, the resulting large gradients in velocity, species concentrations, and temperature create conditions prone to the development of Kelvin-Helmholtz, centrifugal, and Rayleigh-Taylor instabilities \cite{sirignano_turbine_2012}. These instabilities can significantly impact energy conversion, heat transfer, and force loading on the turbine blades.

Computational investigations of turbulent, accelerating, turning, reacting flows relevant to the turbine-burner concept are extensive, yet further studies in realistic configurations are still needed to advance its technology readiness. Sirignano and Kim \cite{sirignano_diffusion_1997} developed a similarity solution for a laminar compressible reacting mixing layer under favorable pressure gradient. Fang et al. \cite{fang_ignition_2001} extended this to non-similar boundary-layer formulations, and Mehring et al. \cite{mehring_ignition_2001} incorporated turbulence using Reynolds-Averaged Navier-Stokes (RANS) equations. Cai et al. \cite{cai_combustion_2001,cai_ignition_2001} solved the full compressible two-dimensional Navier–Stokes equations for turbulent transonic reacting flows in curved ducts, representative of turbine passages with both streamwise and transverse acceleration. Cheng et al. \cite{cheng_nonpremixed_2007,cheng_nonpremixed_2008,cheng_reacting_2009} studied non-premixed combustion in similar geometries, capturing laminar-to-turbulent transition and transonic regimes with multi-species chemistry. Zhu et al. \cite{zhu_numerical_2024,zhu_large-eddy_2025} extended these efforts to RANS and Large Eddy Simulations (LES) of realistic turbine stator–rotor configurations with vitiated-air oxidation, accounting for blade rotation.

All prior works employed simplified one-step kinetic (OSK) models, which poorly represent combustion heat release. In contrast, flamelet approaches \cite{bilger_structure_1976,peters_local_1983,peters_laminar_1984,peters_laminar_1988} treat turbulent diffusion flames as ensembles of laminar counterflow flamelets at the Kolmogorov scale, precomputed offline and stored in tabulated form. This strategy avoids solving species transport equations at runtime while enabling finite-rate chemistry through complex multi-step mechanisms \cite{peters_laminar_1984}. The ability to work with finite-rate chemistry potentially allows for a more accurate capture of the combustion heat release compared to simplified combustion models such as OSK approaches.

Walsh et al. \cite{walsh_turbulent_2025} extended the two-dimensional RANS boundary-layer approximation mixing layer studies of Fang et al. and Mehring et al. \cite{fang_ignition_2001, mehring_ignition_2001} by incorporating a methane 13-species Flamelet Progress Variable (FPV) approach \cite{pierce_progress_2001,pierce_progress-variable_2004} for the simulation of methane/vitiated-air combustion with detailed finite rate chemistry. Using the FPV approach, they observed lower peak flame temperatures, shorter flame stand-offs (indicating faster kinetics), and different mixing layer growth. 

In a subsequent study, Walsh et al. \cite{walsh_performance_2026} showed that the decoupling between resolved-scale and flamelet-scale strain rates in the FPV formulation leads to the preferential selection of equilibrium flamelet solutions in regions of high strain, resulting in nonphysical heat release rates. To address this, they proposed a novel compressible, mixture-fraction-based flamelet model that uses the turbulent kinetic energy dissipation rate, $\epsilon$, as the flamelet tracking parameter instead of the progress variable. This approach builds on the theoretical framework of Sirignano et al. \cite{sirignano_flamelet_2024}, who proposed a scaling methodology based on $\epsilon$ obtained from RANS or LES simulations. Their analysis relates $\epsilon$ to the local turbulent kinetic energy and the viscous dissipation rate at the Kolmogorov scale, enabling estimation of the smallest eddy turnover times and associated strain rates. Through this framework, the strain rate imposed on the flamelet (representing the mechanical constraint governing its structure) can be determined consistently with the local turbulence cascade. 

The present paper has two primary objectives. First, it builds upon the study of Walsh et al. \cite{walsh_turbulent_2025} by (1) considering JP-5 as a more practical fuel, and (2) incorporating a realistic turbine-passage computational configuration. Second, it aims to demonstrate the performance of the novel compressible flamelet model with $\epsilon$-based tracking \cite{walsh_performance_2026} in a practical high-speed flow of engineering relevance.

In order to pursue these objectives, two-dimensional RANS computations of the reacting flow in a turbine stator passage are conducted. In the absence of experimental data for JP-5 combustion in a turbine stator passage, the $\epsilon$-based computations are given partial validation by comparison with methane/vitiated-air simulations. Both the traditional OSK approach and the proposed $\epsilon$-based model are applied to the methane/vitiated-air case so that the effects of using a flamelet-based formulation instead of the OSK model can be isolated from the influence of the fuel chemistry.

\section{Governing Equations and Combustion Models}
The governing equations required for the compressible, multi-species, reacting RANS computations using both the OSK and the $\epsilon$-based flamelet models are provided in this section.
\subsection{Governing Equations}
\subsubsection{Reynolds-Averaged Navier-Stokes Equations}
The Reynolds-Averaged Navier-Stokes (RANS) equations for compressible flows with $N$ individual species are expressed by the following transport equations for mass, momentum and energy
\begin{subequations} \label{eq:navier-stokes}
    \begin{align}
        \frac{\partial\bar{\rho}}{\partial t} + \nabla \cdot (\bar{\rho}\widetilde{\mathbf{V}}) & 
            = 0 \label{eq:continuity} \\
        \frac{\partial(\bar{\rho} \widetilde{\mathbf{V}})}{\partial t} + \nabla \cdot (\bar{\rho}\widetilde{\mathbf{V}} \widetilde{\mathbf{V}}) &
            = -\nabla \bar{p} + \nabla \cdot \boldsymbol{\tau} \label{eq:momentum} \\
        \frac{\partial(\bar{\rho} \widetilde{E})}{\partial t} + \nabla \cdot (\bar{\rho} \widetilde{E}\widetilde{\mathbf{V}}) & = -\nabla \cdot (\bar{p} \widetilde{\mathbf{V}}) + \nabla \cdot (\widetilde{\mathbf{V}} \cdot \boldsymbol{\tau}) - \nabla \cdot \mathbf{q} + \widetilde{\dot Q} \label{eq:energy} 
    \end{align}
\end{subequations}

Here $\widetilde{E}$ is the total sensible energy given by
\begin{equation}
    \widetilde{E} = \tilde{h} - \frac{\bar{p}}{\bar{\rho}} + \frac{1}{2}\widetilde{\mathbf{V}} \cdot \widetilde{\mathbf{V}}
\end{equation}
with
\begin{equation}
    \tilde{h} = \sum_{n=1}^{N}{\widetilde{Y}_n \tilde{h}_n}, \ \ \tilde{h}_n = \int_{T_{\mathrm{ref}}}^{\widetilde{T}}{C_{p,n}(T)\mathrm{d}T}
\end{equation}
where $C_{p,n}$ is given by NASA polynomials \cite{mcbride_coefficients_1993} and $\widetilde{Y}_n$ is the mass fraction of species $n$. The heat release rate $\widetilde{\dot Q}$ in Eq. (\ref{eq:energy}) is provided by the combustion model and will be discussed later. 

The perfect gas equation of state is assumed,
\begin{equation}\label{eq:eos}
    \bar{p} = \bar{\rho}R\widetilde{T} \:,
\end{equation}
where $R$ is the gas constant of the mixture, computed by the mass-weighted summation of the gas constant of each species $R_n$ with $R_n = R_0/W_n$.

\subsubsection{Species Transport and Flamelet Model Transport Equations}
For the OSK and the $\epsilon$-based flamelet combustion models, species transport is solved through $N$ partial density transport equations:
\begin{equation} \label{eq:species}
    \frac{\partial{\bar{\rho} \widetilde{Y}_n}}{\partial t} + \nabla \cdot (\bar{\rho} \widetilde{Y}_n \widetilde{\mathbf{V}}) = -\nabla \cdot \mathbf{j}_n + \widetilde{\dot{\omega}}_n, \ n = 1, \ 2, \ ..., \ N  \;,
\end{equation}
where $\widetilde{\dot{\omega}}_n$ are the species production rates. In the case of OSK, these chemical rates are provided by a single reaction based on resolved-scale concentrations. In the case of the $\epsilon$-based flamelet model, the rates are provided by the precomputed flamelet libraries. Coupling between these libraries and the resolved flow field is achieved through the transport of the mean mixture fraction $\widetilde{Z}$ and its mean variance $\widetilde{Z^{''2}}$ . These quantities are governed by their respective transport equations:

\begin{equation}
\frac{\partial \bar{\rho} \widetilde{Z}}{\partial t} + \nabla\cdot(\bar{\rho} \widetilde{Z}\widetilde{\mathbf{V}}) = -\nabla\cdot\mathbf{j}_z
\label{zequation}
\end{equation}

\begin{equation}
\frac{\partial \bar{\rho} \widetilde{Z^{''2}}}{\partial t} + \nabla\cdot(\bar{\rho} \widetilde{Z^{''2}}\widetilde{\mathbf{V}}) = -\nabla\cdot\mathbf{j}_{z^{''2}}+
2\frac{\mu_{T}}{\mathrm{Sc}_{T}}\nabla\widetilde{Z}\cdot\nabla\widetilde{Z}-\bar{\rho}\widetilde{\chi}
\label{varzequation}
\end{equation}
where the final term in the variance equation represents the mean scalar dissipation rate. This quantity is modeled by relating the integral scalar time-scale and the turbulent flow time-scale, yielding an expression proportional to the turbulent dissipation rate and the mixture-fraction variance:
\begin{equation}
    \widetilde{\chi}=C_{\chi}\frac{\epsilon}{k}\widetilde{Z''^2}\hspace{1mm},
\end{equation}
where $k$ and $\epsilon$ are the turbulent kinetic energy and turbulent dissipation rate, respectively \cite{peters_turbulent_2000}. The constant $C_{\chi}$ is set to $C_{\chi}=2.0$ \cite{janicka_two-variables_1979}. 

\subsubsection{Transport Properties}
The transport terms in the previous equations are given by 
\begin{subequations}
    \begin{align}
        \boldsymbol{\tau} &= 2(\mu + \mu_T) \left[ \mathbf{S} - \frac{1}{3}(\nabla \cdot \widetilde{\mathbf{V}} ) \mathbf{I} \right], \ \mathbf{S} = \frac{1}{2} \left[\nabla \widetilde{\mathbf{V}} + (\nabla \widetilde{\mathbf{V}})^T \right] \\
        \mathbf{j}_n &= -\left( \frac{\mu}{\mathrm{Sc}_n} + \frac{\mu_T}{\mathrm{Sc}_T} \right) \nabla \widetilde{Y}_n \\
        \mathbf{q} &= -\left( \frac{\mu}{\mathrm{Pr}} + \frac{\mu_T}{\mathrm{Pr}_T} \right) \left( \nabla \tilde{h} - \sum_{n=1}^{N}{\tilde{h}_n \nabla \widetilde{Y}_n} \right) + \sum_{n=1}^{N}{\tilde{h}_n\mathbf{j}_n} \label{eq:heat_flux} \\ 
        \mathbf{j}_z &= -\left(\frac{\mu}{\mathrm{Sc}}+\frac{\mu_{T}}{\mathrm{Sc}_{T}}\right)\nabla\widetilde{Z} \\
        \mathbf{j}_{z^{''2}} &= -\left(\frac{\mu}{\mathrm{Sc}}+\frac{\mu_{T}}{\mathrm{Sc}_{T}}\right)\nabla\widetilde{Z^{''2}}
    \end{align}
\end{subequations}
The molecular viscosity, $\mu$, is computed by the mass-weighted summation of the molecular viscosity of each species given by the Sutherland law. All species Schmidt numbers, $\mathrm{Sc}_n$, are assumed to be 1.0. The turbulent Schmidt number $\mathrm{Sc}_T$, the Prandtl number $\mathrm{Pr}$, and the turbulent Prandtl number $\mathrm{Pr}_T$ are all set as 0.7 in the present study. The last term in Eq. (\ref{eq:heat_flux}) accounts for the energy transport due to mass diffusion of each species with different enthalpy.

\subsubsection{Turbulence Model}
The turbulent viscosity $\mu_T$ is determined by the $k\mbox{-}\omega$ Shear-Stress Transport (SST) model presented by Menter et al. \cite{menter_ten_2003} in 2003: 
\begin{subequations} \label{eq:k-omega}
    \begin{align}
        \dfrac{\partial \bar{\rho} k}{\partial t} + \nabla\cdot(\bar{\rho} k \widetilde{\mathbf{V}}) &= P - \beta^*\bar{\rho} k\omega + \nabla \cdot [(\mu+\sigma_k\mu_T) \nabla k] \\
        \dfrac{\partial \bar{\rho} \omega}{\partial t} + \nabla\cdot(\bar{\rho}\omega \widetilde{\mathbf{V}}) &= \dfrac{\gamma\bar{\rho}}{\mu_T}P - \beta\bar{\rho}\omega^2 + \nabla\cdot\left[(\mu+\sigma_\omega\mu_T)\nabla\omega\right] + 2(1-F_1)\dfrac{\bar{\rho}\sigma_{\omega2}}{\omega}\nabla k\cdot \nabla\omega
    \end{align}
\end{subequations}

The production term $P$ is defined as
\begin{equation} \label{eq:turbulent_Pk}
    P = \mathrm{min}(\mu_TS^2, 10\beta^*\bar{\rho} k\omega)
\end{equation}
where $S = \sqrt{2\mathbf{S}:\mathbf{S}}$ is the magnitude of the strain-rate tensor $\mathbf{S}$.
The turbulent viscosity is then computed by 
\begin{equation} 
    \mu_T = \dfrac{a_1\bar{\rho} k}{\mathrm{max}\left(a_1 \omega, F_2S\right)}
\end{equation}

The definitions of the blending functions $F_1$ and $F_2$, as well as the model constants can be found in Ref. \cite{menter_ten_2003}.

\subsection{One-Step Kinetics Combustion Model}
A resolved-scale one-step kinetics combustion model for $\mathrm{CH_4}$/vitiated-air combustion is used as a benchmark case. The global reaction in this case is given by 
\begin{equation} \label{eq:methane_reaction}
    {\rm CH}_4 + 2 {\rm O}_2 + 7.52 {\rm N}_2 \rightarrow {\rm CO}_2 + 2 {\rm H}_2{\rm O} + 7.52 {\rm N}_2 \:.
\end{equation}
Chemical reaction rates are determined from resolved-scale species concentrations using the following modified Arrhenius expression
\begin{equation} \label{eq:Arrhenius_expression}
    {\widetilde{\varepsilon}} = A \widetilde{T}^\beta e^{-E_a/(R_0 \widetilde{T})} \widetilde{C}_{\rm{CH_4}}^a \widetilde{C}_{\rm{O_2}}^b 
\end{equation}
Here the concentrations are given by $\widetilde{C}_n = \bar{\rho} \widetilde{Y}_n/W_n$. The constants $A = 1.3 \times 10^{9} \, \rm{s^{-1}}$, $\beta = 0$, $E_a = 202.506 \, \rm{kJ/mol}$, $a = -0.3$, and $b = 1.3$ specific to this methane reaction are given by Westbrook and Dryer \cite{westbrook_chemical_1984}. A total of five species mass fraction are considered in this model, including $\mathrm{CH_4}$, $\mathrm{O_2}$, $\mathrm{N_2}$, $\mathrm{CO_2}$ and $\mathrm{H_2O}$. The species production rates are determined using
\begin{equation}
    \widetilde{\dot\omega}_n = W_n(v_n^{\prime\prime} - v_n^\prime) \widetilde{\varepsilon}
\end{equation}
where $v_n^\prime$ is the stoichiometric coefficient for reactant $n$ and $v_n^{\prime\prime}$ is the stoichiometric coefficient for product $n$ as determined by Eq. (\ref{eq:methane_reaction}).

The mean heat release rate term $\widetilde{\dot Q}$ appearing on the right-hand side of Eq. (\ref{eq:energy}) is defined using the resolved-scale species production rates,
\begin{equation}
    \widetilde{\dot Q} = -\sum_{n=1}^{N}{\widetilde{\dot\omega}_n h^0_{f,n}} \:.
\end{equation}

\subsection{Epsilon-based Flamelet Combustion Model}\label{sec:combustion_model}
In this work, we employ the $\epsilon$-based flamelet combustion model introduced by Walsh et al.~\cite{walsh_performance_2026}, which infers the local strain rate imposed on the flamelet from the turbulent kinetic energy dissipation rate, $\epsilon$, following the formulation proposed by Sirignano et al.~\cite{sirignano_flamelet_2024}.

Specifically, flamelet libraries are generated for the resolved-scale inlet reactant compositions and temperatures by solving the system of steady flamelet equations in mixture fraction space under a counterflow configuration~\cite{peters_laminar_1984}, namely,
\begin{equation}\label{eq:steady_flamelet}
-\rho \frac{\chi(Z)}{2}\frac{\partial^2 \psi_j}{\partial Z^2}=\dot{\omega}_j, \hspace{1mm}j=1,2,...,M+1
\end{equation}
for varying stoichiometric scalar dissipation rates $\chi_{st} = \chi(Z_{st})$ and background pressures, $p$. Here, $Z$ denotes the mixture fraction and $\psi_j$ represents the reactive scalars, consisting of the mass fractions of the $M$ species considered in the reaction mechanism and temperature. The source terms $\dot{\omega}_j$ correspond to species reaction rates for the mass fraction equations and to the heat release rate in the energy equation.

The scalar dissipation rate $\chi(Z)$ is assumed to follow the standard form,
\begin{equation}\label{eq:chi_form}
\chi(Z)=\frac{2S^*}{\pi}\exp{(2\mathrm{erfc}^{-1}(2Z)^2)} \:,
\end{equation}
which removes the need to solve the momentum equations. While this approach is widely used in the literature \cite{peters_turbulent_2000}, more comprehensive flamelet formulations that additionally solve the momentum equations have been proposed \cite{sirignano_three-dimensional_2022,sirignano_inward_2022,hellwig_vortex_2025,hellwig_three-dimensional_2025}. In this expression, $S^*$ is the imposed strain rate at the counterflow inlet. Notably, $\chi(Z)$ is uniquely parameterized by $S^*$, which is proportional to the stoichiometric scalar dissipation rate $\chi_{st}$, such that the solutions of Eq. (\ref{eq:steady_flamelet}) may be parametrized by $S^*$:
\begin{equation}\label{eq:psi_sstar}
\psi_j = \psi_j(Z,S^*,p)
\end{equation}
where $p$ denotes the background pressure. The inclusion of pressure as a third dimension is justified by the strong spatial variation of pressure within the turbine stator, allowing the model to capture the effect of local pressure on the chemical kinetics.

In the present work, libraries of the form of Eq.~(\ref{eq:psi_sstar}) are generated for both the $\mathrm{CH_4}$/vitiated-air and JP-5/vitiated-air cases. The $\mathrm{CH_4}$/vitiated-air case is included for comparison with the OSK $\mathrm{CH_4}$ combustion model. For both cases, the FlameMaster code~\cite{flamemaster} is used to solve the flamelet equations, Eq.~(\ref{eq:steady_flamelet}), under the following boundary conditions: hot vitiated air at 1650 K on the air side with a composition of 73.77\% $\mathrm{N_2}$, 11.01\% $\mathrm{O_2}$, 8.04\% $\mathrm{CO_2}$, and 7.18\% $\mathrm{H_2O}$ by mass; and either pure $\mathrm{CH_4}$ or pure JP-5 at 400 K. These boundary conditions coincide with the resolved-scale turbine inlet stream conditions. The composition and temperature of the vitiated air are determined assuming upstream JP-5/air combustion with a fuel-to-air ratio of 0.03\% in the engine’s primary combustor.
For the $\mathrm{CH_4}$ case, the global reaction represented by Eq.~(\ref{eq:methane_reaction}) is modeled using a 13-species ($M=13$), 32-reaction skeletal reduction of Version 1.0 of the Foundational Fuel Chemistry Model (FFCM-1)~\cite{smith_foundational_2016,tao_critical_2018}, previously employed in both flamelet progress-variable computations~\cite{zhan_combustion_2024,walsh_turbulent_2025} and $\epsilon$-based flamelet models~\cite{walsh_performance_2026}. The species included in the skeletal reduction are $\mathrm{H_2}$, $\mathrm{H}$, $\mathrm{O_2}$, $\mathrm{O}$, $\mathrm{OH}$, $\mathrm{HO_2}$, $\mathrm{H_2O}$, $\mathrm{CH_3}$, $\mathrm{CH_4}$, $\mathrm{CO}$, $\mathrm{CO_2}$, $\mathrm{CH_2O}$, and $\mathrm{N_2}$. Nitrogen is treated as inert, participating in elementary reactions only as a third body.

\begin{figure}
    \centering
    \includegraphics[width=0.4\linewidth]{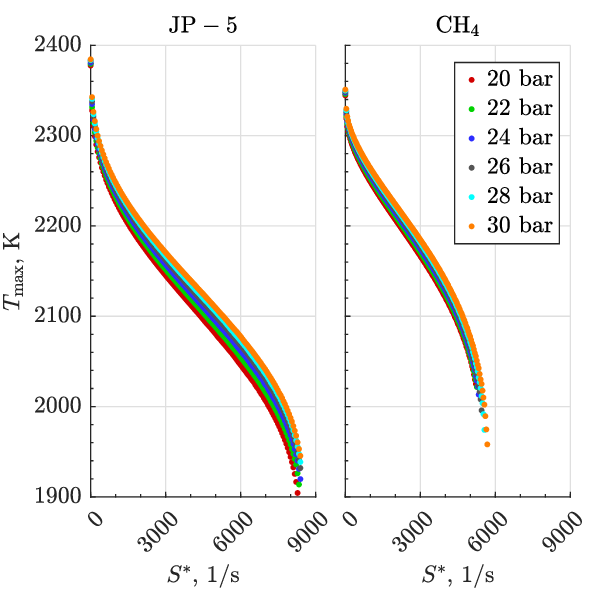}
    \caption{Maximum flamelet temperatures in terms of $\mathbf{S^*}$.}
    \label{fig:sscurves}
\end{figure}

For the JP-5 case, the global reaction is represented as 
\begin{equation}
\mathrm{C_{12}H_{23} + 18.5O_2 + 69.56N_2 \rightarrow 12CO_2 + 11.5H_2O + 69.56N_2},
\end{equation}
where $\mathrm{C_{12}H_{23}}$ denotes the modeled surrogate hydrocarbon representing JP-5. The HyChem high-temperature A3 reaction mechanism \cite{wang_physics-based_2018,xu_physics-based_2018}, which includes 119 chemical species ($M=119$) and 841 elementary reactions, is employed for this case. This mechanism includes pyrolysis pathways in which, at sufficiently high temperatures, the large JP-5 hydrocarbons decompose endothermically into smaller hydrocarbons that subsequently participate in oxidation reactions once $\mathrm{O_2}$ becomes available. As in the $\mathrm{CH_4}$ mechanism, $\mathrm{N_2}$ is treated as inert.

Figure~\ref{fig:sscurves} shows the flamelet solutions for both the $\mathrm{CH_4}$/vitiated-air and JP-5/vitiated-air cases. The maximum flamelet temperature (in terms of Z) is plotted as a function of the inflow strain rate, $S^*$, for different pressures spanning the range expected at the resolved scale. These relations correspond to the well-known S-shaped curves; however, only the stable branch, from $S^*$ near zero up to the flammability limit (maximum attainable strain rate), is considered. The unstable branches are excluded in the $\epsilon$-based approach, as they are not expected to occur with significant frequency. Beyond the flammability limit, the flamelet is quenched. Notably, the JP-5 flamelets exhibit a flammability limit approximately two thirds higher than that of $\mathrm{CH_4}$, which has implications for the resolved-scale combustion behavior.

Mean reactive scalars are determined from Eq. (\ref{eq:psi_sstar}) using the presumed-shape Probability Density Function (PDF) approach, which is widely employed in flamelet-based combustion modeling (see Refs.~\cite{peters_turbulent_2000,pierce_progress-variable_2004,pecnik_reynolds-averaged_2012,saghafian_efficient_2015,nguyen_impacts_2018,nguyen_spontaneous_2019,shadram_neural_2021,shadram_physics-aware_2022,zhan_combustion_2024,walsh_performance_2026}). The Favre-averaged reactive scalars are obtained through the convolutions,
\begin{equation}\label{eq:psi_mean}
\widetilde{\psi}_j(\widetilde{Z},\widetilde{Z''^2},S^*,\bar{p})=\int_0^1 \psi_j(Z,S^*,\bar{p})\widetilde{P}(Z,\widetilde{Z},\widetilde{Z''^2})dZ \:.
\end{equation}

Here, we employ the standard $\beta$-PDF for the mixture fraction $Z$, while for $\epsilon$ a Dirac $\delta$-PDF is used, reflecting the current absence of a more appropriate statistical description. This assumption may warrant further investigation. A $\delta$-PDF is also applied to the pressure. For further details on the convolution procedure used here, see Refs. \cite{walsh_turbulent_2025} and \cite{walsh_performance_2026}.

These precomputed mean flamelet solutions are then stored in library form for the CFD to access at runtime, during which, $\widetilde{Z}$ and $\widetilde{Z''^2}$ are provided by their respective resolved-scale transport equations, Eq. (\ref{zequation}) and Eq. (\ref{varzequation}), respectively. The pressure $\bar{p}$ is provided by the equation of state defined in Eq. (\ref{eq:eos}). All that remains is to determine the flamelet inflow strain rate, $S^*$. This is accomplished by employing the turbulent kinetic energy dissipation rate, $\epsilon$, to infer $S^*$, following the theoretical framework proposed by Sirignano et al.~\cite{sirignano_flamelet_2024}. In their formulation, $S^*$, is related to $\epsilon$ through a gradient-based scaling argument:
\begin{eqnarray}\label{eq:sstar-eps}
S^* = \frac{1}{2}\sqrt{ \frac{C_{vd}\ \epsilon}{ \nu[S_1^2 + 1 - S_1]} }
\end{eqnarray}

\noindent where $C_{vd}$ is a dimensionless coefficient accounting for the distribution of viscous dissipation across a range of the smallest turbulent scales. For the counterflow solution to be physically valid, the constraint $C_{vd} < 1$ must be satisfied. While this coefficient remains to be definitively determined from direct numerical simulation (DNS) data, a value of $C_{vd} = 1$ is adopted in the present study, consistent with the assumptions outlined in \cite{sirignano_flamelet_2024}. The parameter $S_1 = 1/2$ corresponds to the axisymmetric counterflow configuration considered here, and $\nu$ is the molecular kinematic viscosity, obtained from the resolved-scale transport properties. Since the SST two-equation turbulence model is employed here, $\epsilon$ is computed from the resolved-scale turbulence quantities $k$ and $\omega$, provided by their transport Eqs. (\ref{eq:k-omega}), using the standard relation
\begin{equation}
    \epsilon = C_{\mu} k\omega \:,
\end{equation}

\noindent where $C_{\mu} = 0.09$ is a model constant specified by the $k-\epsilon$ turbulence model \cite{chien_predictions_1982}. With this formulation, for any given spatial and temporal location in the resolved-scale domain, $\epsilon$ is used to compute the corresponding flamelet-scale strain rate in accordance with turbulence cascade scaling. This approach reflects the well-established physical principle that scalar gradients intensify as the turbulent length scale decreases, enabling a more consistent coupling between turbulence structure and flamelet dynamics. Given that $S^*$ is now known, reactive scalars, $\psi_j$, may be retrieved through quadrilinear interpolation of the flamelet libraries. If the local strain rate $S^*$ exceeds the flammability limit (as described in Fig. \ref{fig:sscurves}), the libraries return $\psi_j$ corresponding to a quenched (non-reacting) solution.

\begin{figure}
\centering
\begin{subfigure}{.475\textwidth}
  \centering
  \includegraphics[width=0.90\textwidth]{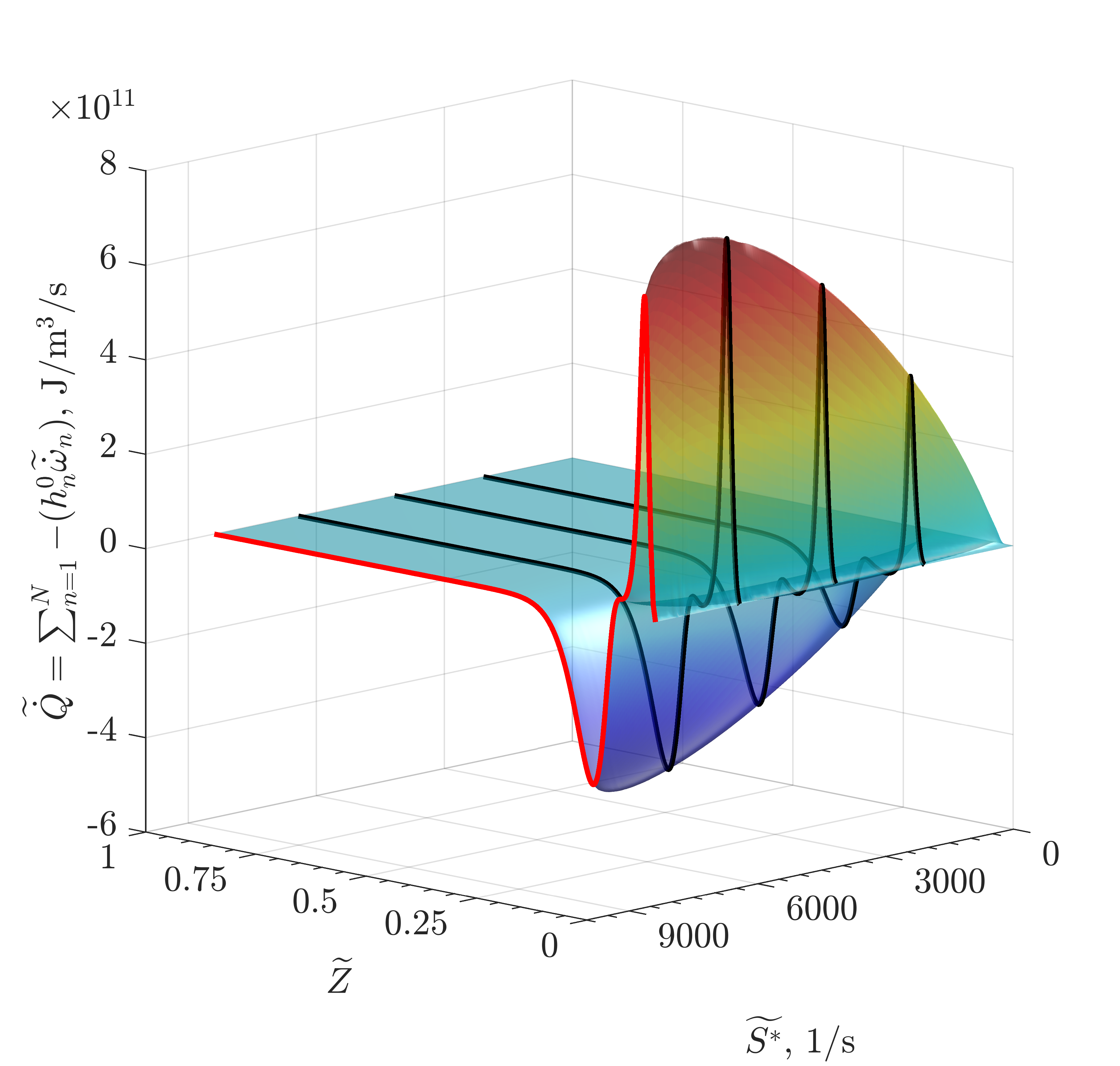}
  \caption{${\widetilde{\dot{Q}}}$ surface. }
  \label{fig:q_jp5_tables}
\end{subfigure}
\begin{subfigure}{.475\textwidth}
  \centering
  \includegraphics[width=1.0\textwidth]{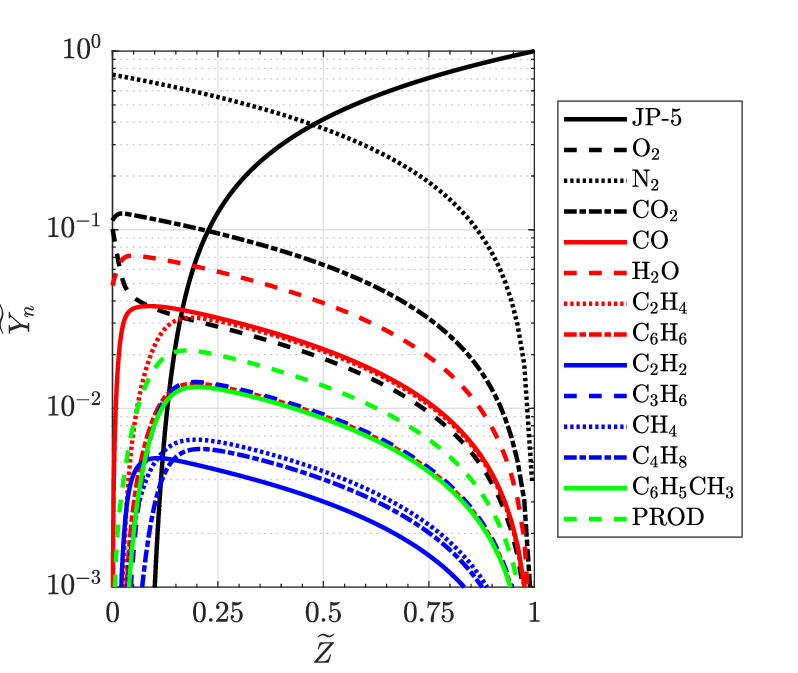}
  \caption{Species mass fractions at the flammability limit $S^*$.}
  \label{fig:yn_tab}
\end{subfigure}%
\caption{Tabulated quantities for the JP-5 case for a $\mathbf{\widetilde{Z''^2}_{\mathrm{norm}}=10^{-8}}$ and $\mathbf{\bar{p}=30}$ bar.}
\label{fig:tables}
\end{figure}

In the $\epsilon$-based approach, species transport is explicitly solved for on the resolved-scale using Eqs. (\ref{eq:species}). Consequentially, what is brought from the flamelet libraries to the resolved scale are the mean species chemical production rates and heat release rate, i.e., $\widetilde{\psi}_{j=1,M} =\widetilde{\dot{\omega}}_j$ and $\widetilde{\psi}_{M+1}=\widetilde{\dot{Q}}$. Under this formulation, a quenched flamelet returns $\widetilde{\dot{\omega}}_j = 0$ for all species and $\widetilde{\dot{Q}}=0$. Figure~\ref{fig:q_jp5_tables} shows the mean heat release rate, $\widetilde{\dot{Q}}$, as a function of $S^*$ and $\widetilde{Z}$ for the JP-5 case, at a normalized mixture-fraction variance of $\widetilde{Z''^2}_{\mathrm{norm}}=10^{-8}$ and a background pressure of $\bar{p}=30$ bar. The black curves on the surface correspond to profiles of $\widetilde{\dot{Q}}$ at constant $S^*$, while the red curve indicates the flammability-limit profile at this pressure. For constant-$S^*$ cuts, a negative trough in $\widetilde{\dot{Q}}$ appears on the fuel side (coming from $\widetilde{Z}=1$), followed by a positive peak. The trough corresponds to endothermic JP-5 pyrolysis, which generates lighter hydrocarbons that subsequently oxidize exothermically, producing the positive peak in $\widetilde{\dot{Q}}$. Beyond the flammability limit in $S^*$, indicated by the red curve, the table returns $\widetilde{\dot{Q}}=0$.

To reduce the computational cost of solving the full set of $M$ species transport equations on the resolved scale with stiff source terms (note $M=13$ for the $\mathrm{CH_4}$ case and $M=119$ for the JP-5 case), only a reduced subset of major species is explicitly transported on the resolved scale. Let the detailed reaction mechanism contain $M$ species. We solve transport equations for $P+1 = N$ species, with $N < M$, where the subset of $P$ species corresponds to the dominant contributors to the mixture composition. Defining  
\begin{equation}\label{eq:gamma}
\gamma = \sum_{n=1}^{P} \widetilde{Y}_n,
\end{equation}
the remaining fraction is represented by a composite lumped species,  
\begin{equation}\label{eq:prod}
\widetilde{Y}_{N} = 1 - \gamma,
\end{equation}
with a source term ensuring mass conservation,  
\begin{equation}
\widetilde{\dot{\omega}}_{N} = -\sum_{n=1}^{P} \widetilde{\dot{\omega}}_n \:,
\end{equation}

\noindent so that $\sum_{n=1}^{N} \widetilde{\dot{\omega}}_n = 0$. For the $\mathrm{CH_4}$ case, where \( M = 13 \) species, a reduced set of \( P = 6 \) major species including \(\mathrm{O_2}\), \(\mathrm{H_2O}\), \(\mathrm{CH_4}\), \(\mathrm{CO}\), \(\mathrm{CO_2}\), and \(\mathrm{N_2}\) are explicitly tracked, while the remaining minor radicals are lumped into $\widetilde{Y}_N=7$. For the JP-5 case, where $M=119$, a subset of $P=13$ species including JP-5, $\mathrm{O_2}$, $\mathrm{N_2}$, $\mathrm{CO_2}$, $\mathrm{CO}$, $\mathrm{H_2O}$, $\mathrm{C_2H_4}$, $\mathrm{C_6H_6}$, $\mathrm{C_2H_2}$, $\mathrm{C_3H_6}$, $\mathrm{CH_4}$, $\mathrm{C_4H_8}$, $\mathrm{C_6H_5CH_3}$ are explicitly tracked, while all other species considered in the reaction mechanism are lumped into $\widetilde{Y}_{N=14}$. These selections capture at least 95\% and 97\% of the total mass fraction (i.e., \(\gamma \ge 0.95\) and \(\gamma \ge 0.97\)) across the entire flamelet solution spaces for the $\mathrm{CH_4}$ and JP-5, respectively. In this way, we avoid the computational cost of tracking the full set of \( M \) species on the resolved scale, while retaining chemical rates computed with subgrid detailed finite-rate kinetics. Furthermore, the timestep constraint is alleviated since the major species exhibit less stiff source terms than the short-lived radicals. 

Figure~\ref{fig:yn_tab} shows the concentrations of the tracked subset of species ($n = 1,\ldots,P$) as well as the lumped species, labeled “PROD” in the plot, for the JP-5 flamelet model on a logarithmic $y$-axis. The mixture-fraction variance is held constant at a normalized value of $10^{-8}$, and the background pressure is $\bar{p} = 30$ bar. The strain rate $S^*$ corresponds to the flammability-limit value at this pressure. Moving from the fuel side at $\widetilde{Z} = 1$ to the vitiated-air side ($\widetilde{Z} = 0$), JP-5 diffuses toward the oxidizer stream. At sufficiently high temperatures, JP-5 undergoes endothermic pyrolysis, decomposing into smaller hydrocarbons including $\mathrm{C_2H_4}$, $\mathrm{C_6H_6}$, $\mathrm{C_2H_2}$, $\mathrm{C_3H_6}$, $\mathrm{CH_4}$, $\mathrm{C_4H_8}$, and $\mathrm{C_6H_5CH_3}$. The pyrolysis completes as the JP-5 mass fraction rapidly decreases around $\widetilde{Z} = 0.15$, where the smaller hydrocarbon mass fractions reach their peak values. This region corresponds to the endothermic heat-release zone shown in Fig.~\ref{fig:q_jp5_tables}. The resulting lighter hydrocarbons subsequently oxidize with the available $\mathrm{O_2}$ in an exothermic process, during which the hydrocarbons are depleted while $\mathrm{O_2}$ is consumed, leading to the formation of oxidation products including $\mathrm{CO_2}$, $\mathrm{CO}$, and $\mathrm{H_2O}$.

In this approach, the local resolved-scale species concentrations directly influence where chemical reactions can occur. The chemical source terms $\widetilde{\dot{\omega}}_n$ obtained from the flamelet library are therefore scaled according to the availability of reactants on the resolved scale. Specifically, for each species $n$, the source term is corrected as
\begin{equation}
\widetilde{\dot{\omega}}_{n_{\mathrm{corr}}} = \alpha\widetilde{\dot{\omega}}_n, \qquad n = 1, 2, \ldots, N,
\end{equation}
where the scalar coefficient $\alpha$ is chosen independently at each computational cell as the minimum value that ensures positivity of all species mass fractions. This correction prevents unphysical behavior that could occur when a flamelet, evaluated at a given $\widetilde{Z}$, predicts finite or negative production rates for reactants whose resolved-scale mass fractions are zero or vanishingly small. In effect, the scaling enforces that no chemical source term is applied in regions devoid of reactants, thereby ensuring a physically consistent coupling between the flamelet chemistry and the resolved species fields.

\section{Computational Setup}
\subsection{Flow Configuration and Grid}\label{sec:flow_conf}
\begin{figure}[h]
    \centering
    \includegraphics[width=0.5\linewidth]{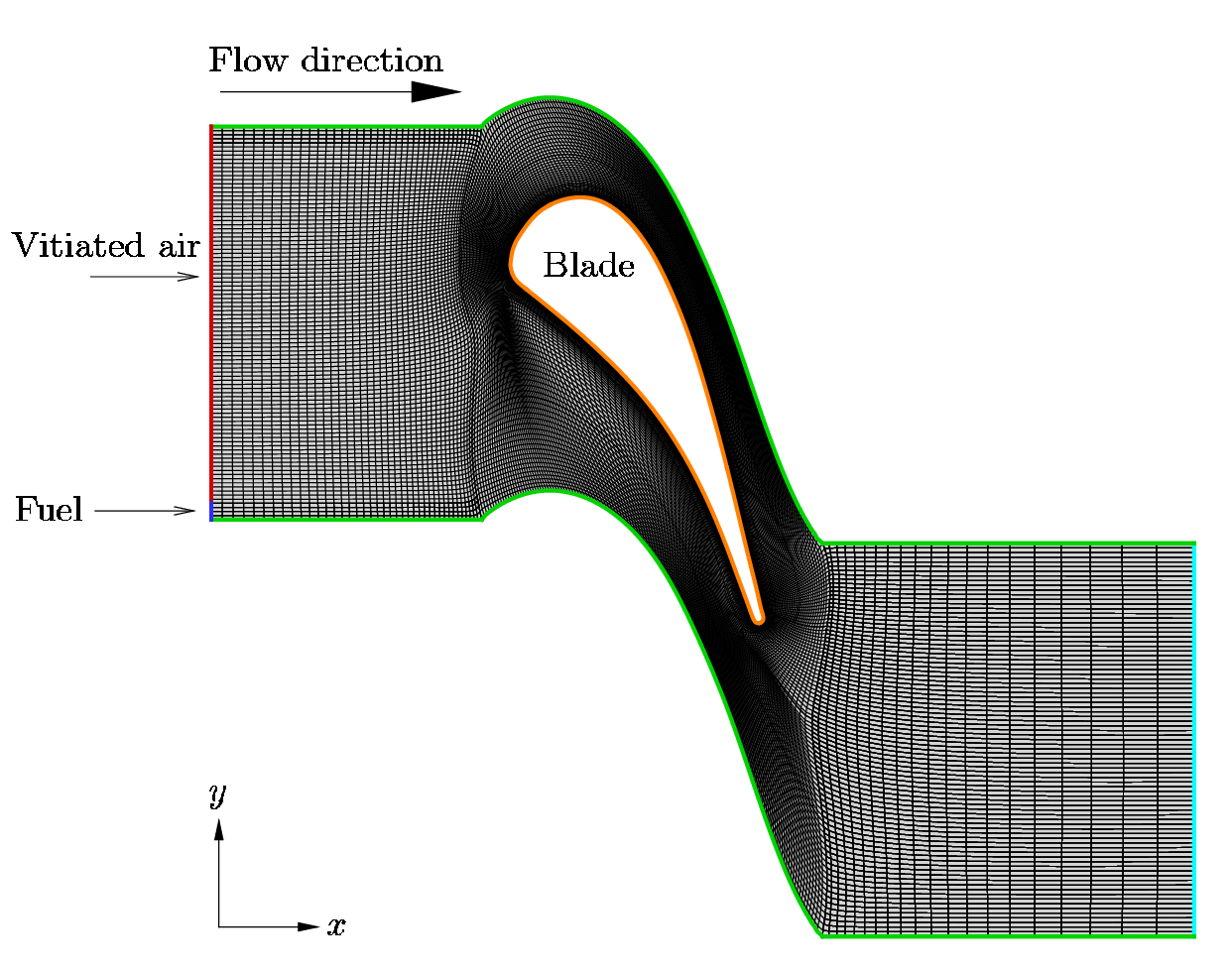}
    \caption{Flow configuration and grid.}
    \label{fig:grid}
\end{figure}

The reacting flow in  the highly loaded transonic \textit{VKI LS89} turbine nozzle vane \cite{arts_aero-thermal_1992} is simulated and compared for cases with JP-5 and $\mathrm{CH_4}$ inlets. The chord of the vane is 76.674 mm, and the pitch-to-chord ratio is 0.85. The stagger angle is 55$^\circ$ from the axial direction. The multiblock structured grid shown in Fig. \ref{fig:grid} is generated for a single cascade passage with translational periodicity on the pitch-wise boundaries (corresponding to the green curves in Fig. \ref{fig:grid}). 317 grid points wrap around the blade surface (orange curve in Fig. \ref{fig:grid}), with the points concentrated near the leading edge and trailing edge. The total grid has 26,416 cells, which are divided into 14 blocks for parallel computation. 

The inlet of the turbine cascade is divided into two sections. In the upper section (red line in Fig.~\ref{fig:grid}), vitiated air is injected at a static temperature of 1650~K. In the lower section (dark blue line in Fig.~\ref{fig:grid}), either pure JP-5 vapor or pure $\mathrm{CH_4}$ vapor at a static temperature of 400~K is introduced parallel to the vitiated air, depending on the case. The ratio of vitiated-air to fuel-inlet areas is 23. These inlet compositions and temperatures correspond to the boundary conditions of the counterflow configurations used to solve the flamelet equations in Eq.~(\ref{eq:steady_flamelet}). Zero flow angle with respect to the x axis is specified for both streams.

The static pressure at both sections of the inlet is set to 30 bar, representing typical turbine inlet static pressures. The total pressure and total temperature at each section are then adjusted to achieve comparable mass-flow rates between the JP-5 and $\mathrm{CH_4}$ configurations. For the JP-5 case, the mass-flow rates of the vitiated-air and fuel streams are 44.32 kg/m/s and 2.68 kg/m/s, respectively. For the $\mathrm{CH_4}$ case, the corresponding values are 40.22 kg/m/s and 2.78 kg/m/s, respectively.

The ratio of the outlet static pressure to the inlet total pressure is set to 0.6, with the outlet location indicated by the blue line in Fig. \ref{fig:grid}. This value is derived from the downstream isentropic Mach number reported in Ref. \cite{arts_aero-thermal_1992}. The resulting static pressure at the outlet is 18 bar, which, together with the corresponding inlet pressure, defines the pressure range employed in the flamelet libraries. Under these conditions, the mean streamwise pressure gradient is approximately 200 atm/m.

A no-slip velocity boundary condition and adiabatic wall boundary conditions are enforced along the blade surfaces.

\subsection{Flow Solver}
An in-house three-dimensional code for simulating steady and unsteady transonic flows for single species within turbomachinery blade rows has been developed, validated, and applied by Refs. \cite{zhu_numerical_2017,zhu_flow_2018,zhu_influence_2018,liu_computational_2025}. The code solves the Navier–Stokes equations together with various turbulence models by using the second-order cell-centered finite-volume method based on a multiblock structured grid. The central schemes with artificial viscosity, flux difference splitting schemes, and advection upstream splitting methods with various options to reconstruct the left and right states have been developed and implemented in  the code. 

Recently, it has been extended to include solving transport equations for multiple species with varying specific heat capacities and appropriate chemistry models, and verified and validated by the two-dimensional steady transonic reacting flows in a mixing layer and a turbine cascade \cite{zhu_numerical_2024}, three-dimensional reacting flow in a turbine stage \cite{zhu_large-eddy_2025} and the three-dimensional unsteady reacting flow in a rocket engine \cite{zhu_simulation_2025}. The convective and viscous fluxes are discretized by the Jameson-Schmidt-Turkel scheme \cite{jameson_numerical_1981} and the second-order central scheme, respectively. The local time-stepping method is introduced to accelerate the convergence to a steady state. Thus, the time $t$ in the governing equations is interpreted as a pseudo-time, and a large enough pseudo-time step determined by the local flowfield can be used in each grid cell since time accuracy is not required for steady-state solutions. An operator-splitting scheme is used to treat the stiff chemical source terms in Eqs. (\ref{eq:energy}) and (\ref{eq:species}) (for details see \cite{zhu_numerical_2024}). Parallel techniques based on the message passing interface (MPI) are adopted to further accelerate the computation by distributing grid blocks among CPU processors.

\section{Results and Discussion}

In this section, computational results for both $\mathrm{CH_4}$ and JP-5 configurations are presented. First, the OSK and $\epsilon$-based flamelet models are compared for the $\mathrm{CH_4}$ configuration to isolate and examine the differences between a resolved-scale simplified-kinetics approach and a subgrid combustion model that incorporates detailed finite-rate chemistry. Subsequently, results for the JP-5 configuration are discussed, representing a more practical fuel with a substantially more complex reaction mechanism. Finally, global performance metrics are discussed.

\subsection{$\boldsymbol{\mathrm{CH_4}}$ Configuration}

\begin{figure}
\centering
\begin{subfigure}{.4\textwidth}
  \centering
  \includegraphics[width=1.0\textwidth]{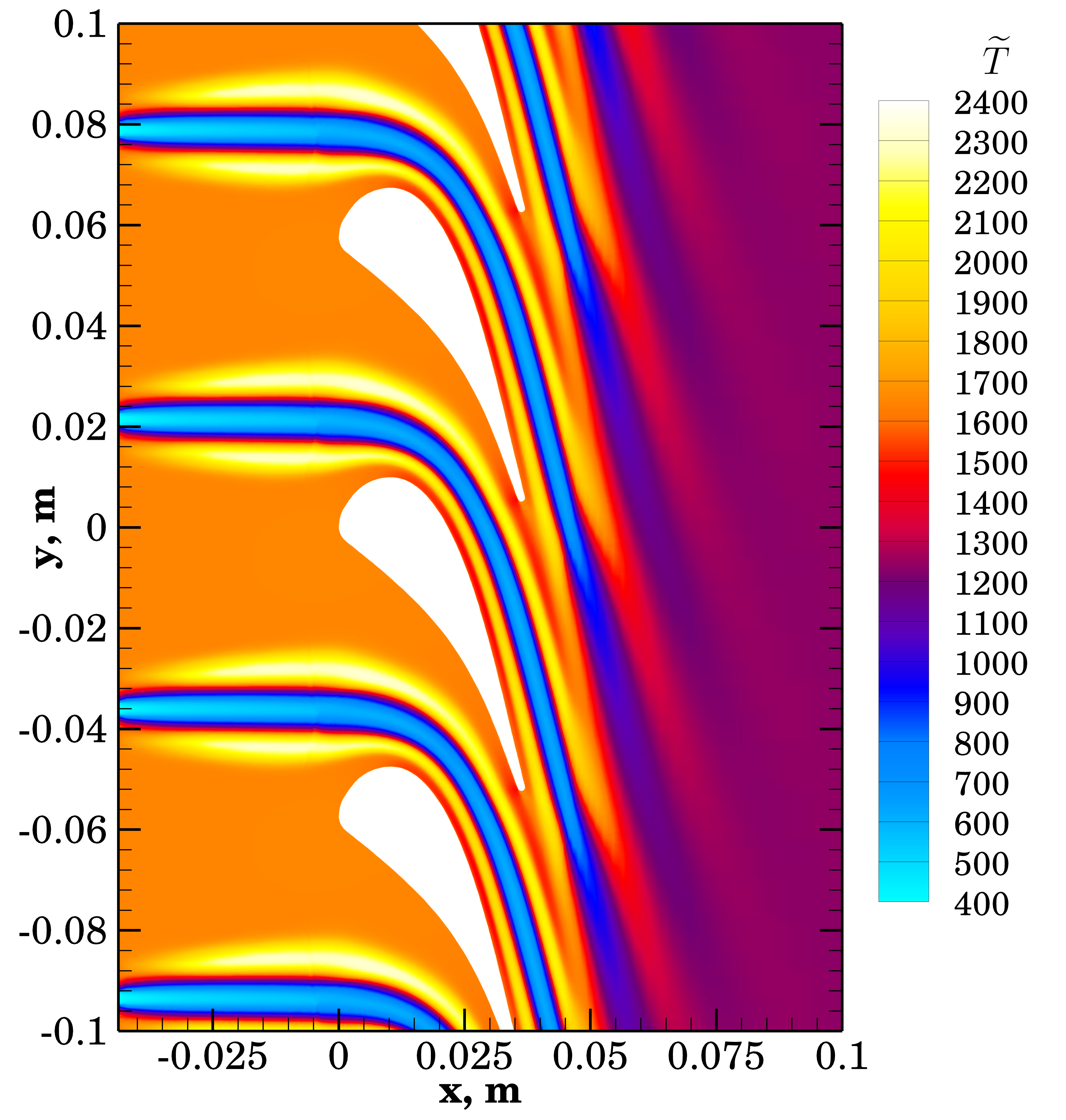}
  \caption{OSK.}
  \label{fig:t_ch4_osk}
\end{subfigure}
\begin{subfigure}{.4\textwidth}
  \centering
  \includegraphics[width=1.0\textwidth]{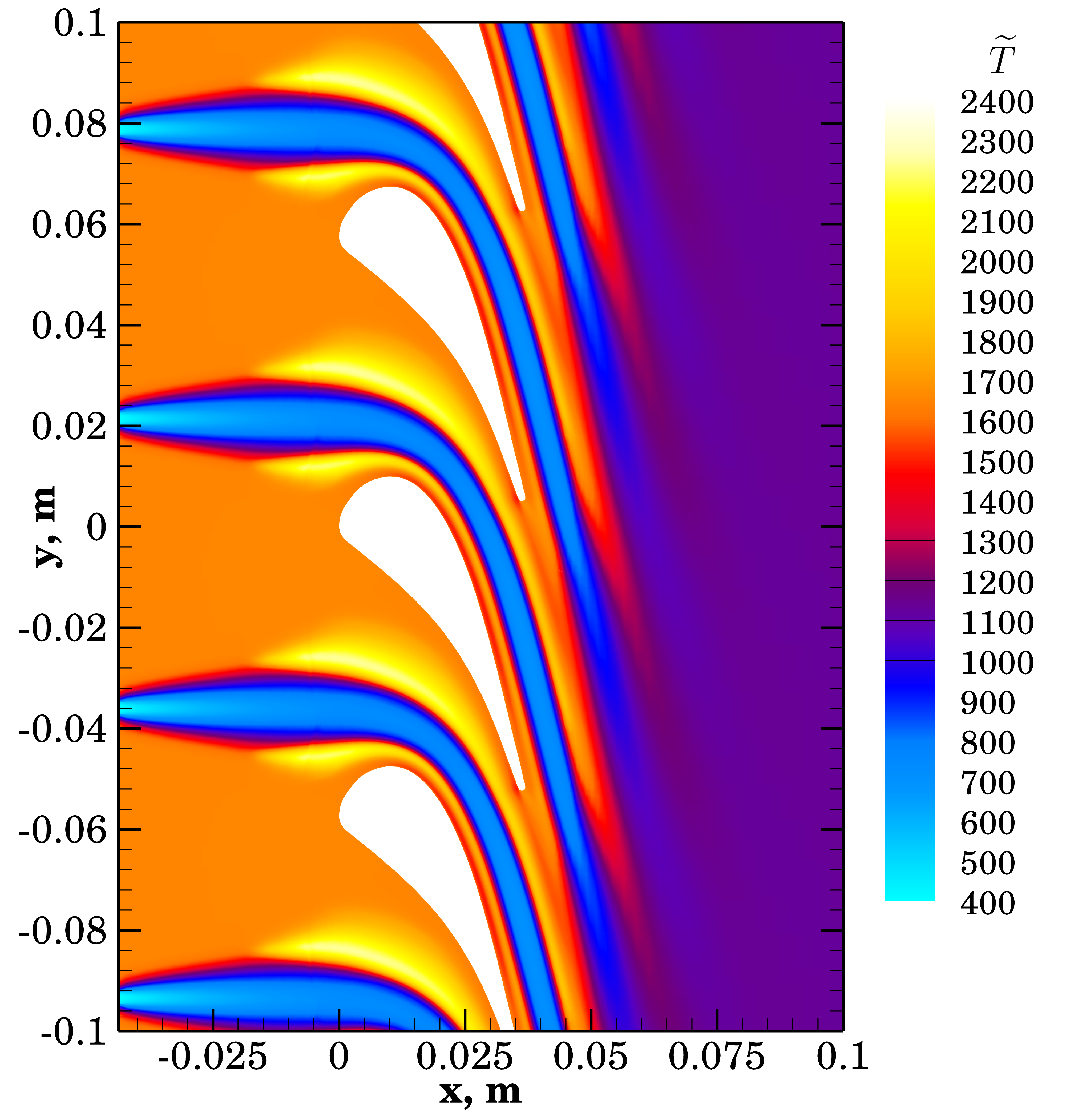}
  \caption{$\boldsymbol{\epsilon}$-based flamelet model.}
  \label{fig:t_ch4_flamelet}
\end{subfigure}%
\caption{$\mathbf{\widetilde{T}}$ fields for the $\boldsymbol{{\mathrm{CH_4}}}$ configuration.}
\label{fig:ch4_t_fields}
\end{figure}

When the two reactant streams enter the computational domain, the differences in velocity, temperature, and composition between them give rise to velocity, thermal, and compositional mixing layers. The mixing of reactants establishes diffusion flames on both sides of the fuel stream, near the stoichiometric mixture fraction. 

Figure \ref{fig:t_ch4_osk} shows the mean temperature contours within the turbine passage obtained using the OSK combustion model. The computational data is stacked vertically and periodically to link the periodic boundaries, thereby providing a clearer visualization of the blade cascade. As described in Section \ref{sec:flow_conf}, however, the computational domain includes only a single blade passage with one fuel and one oxidizer stream. Combustion occurs almost immediately, as indicated by the high-temperature regions in the figure. The cold fuel jet (indicated by the blue colored regions in Fig. \ref{fig:t_ch4_osk}) extends downstream between the blades, with one flame anchoring near the suction surface of the blade, while the other passes through the center of the passage, below the pressure surface. Downstream of the trailing edge, the mid passage flames merge with the suction side flames from the adjacent blades and continue together into the wake region.

Figure~\ref{fig:t_ch4_flamelet} presents the corresponding temperature field obtained using the $\epsilon$-based flamelet combustion model. Peak flame temperatures are approximately 100~K lower than those predicted by the OSK model. This difference is characteristic of comparisons between flamelet and OSK approaches, as the inclusion of detailed finite-rate chemistry in the flamelet model accounts for heat losses due to dissociation. Similar reductions in peak temperature when using flamelet-based models have been reported in Refs.~\cite{walsh_turbulent_2025,walsh_performance_2026}. A distinct flame stand-off distance is observed in the $\epsilon$-based case, with ignition occurring roughly 2.5~cm downstream of the inlet. This delayed ignition enhances shear-driven mixing, as evidenced by the broader development of the cold fuel jet. The increased mixing produces wider reaction zones, as shown by the broader flame regions in Fig.~\ref{fig:t_ch4_flamelet} compared with the OSK results. The overall flame topology remains similar, with one flame anchored near the suction surface and the other located near the passage center. However, the flames exhibit lower strength, manifested by a stronger streamwise temperature decay that prevents them from merging downstream of the trailing edge.

\begin{figure}[h]
\centering
\begin{subfigure}{.4\textwidth}
  \centering
  \includegraphics[width=1.0\textwidth]{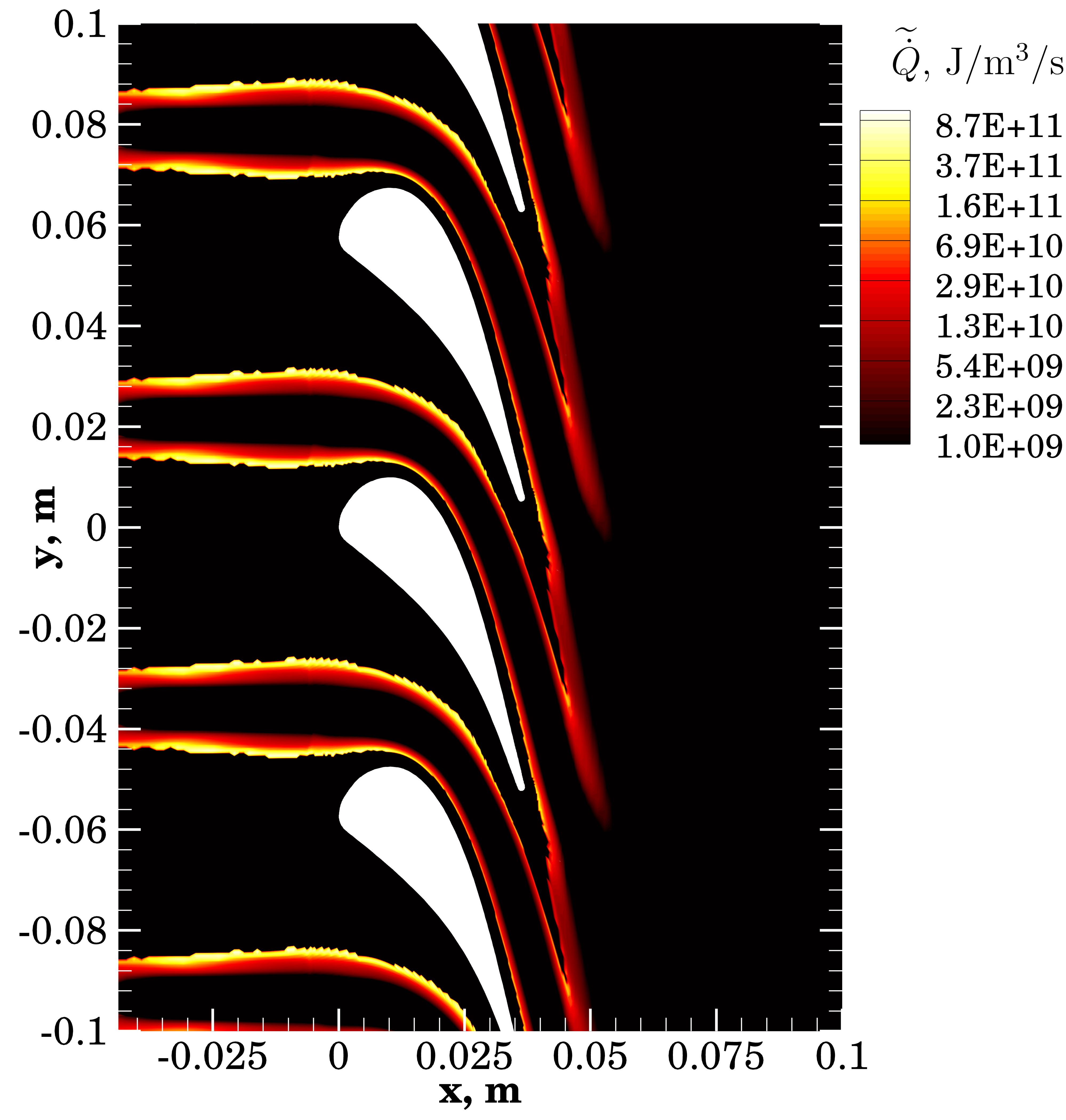}
  \caption{OSK.}
  \label{fig:q_osk}
\end{subfigure}
\begin{subfigure}{.4\textwidth}
  \centering
  \includegraphics[width=1.0\textwidth]{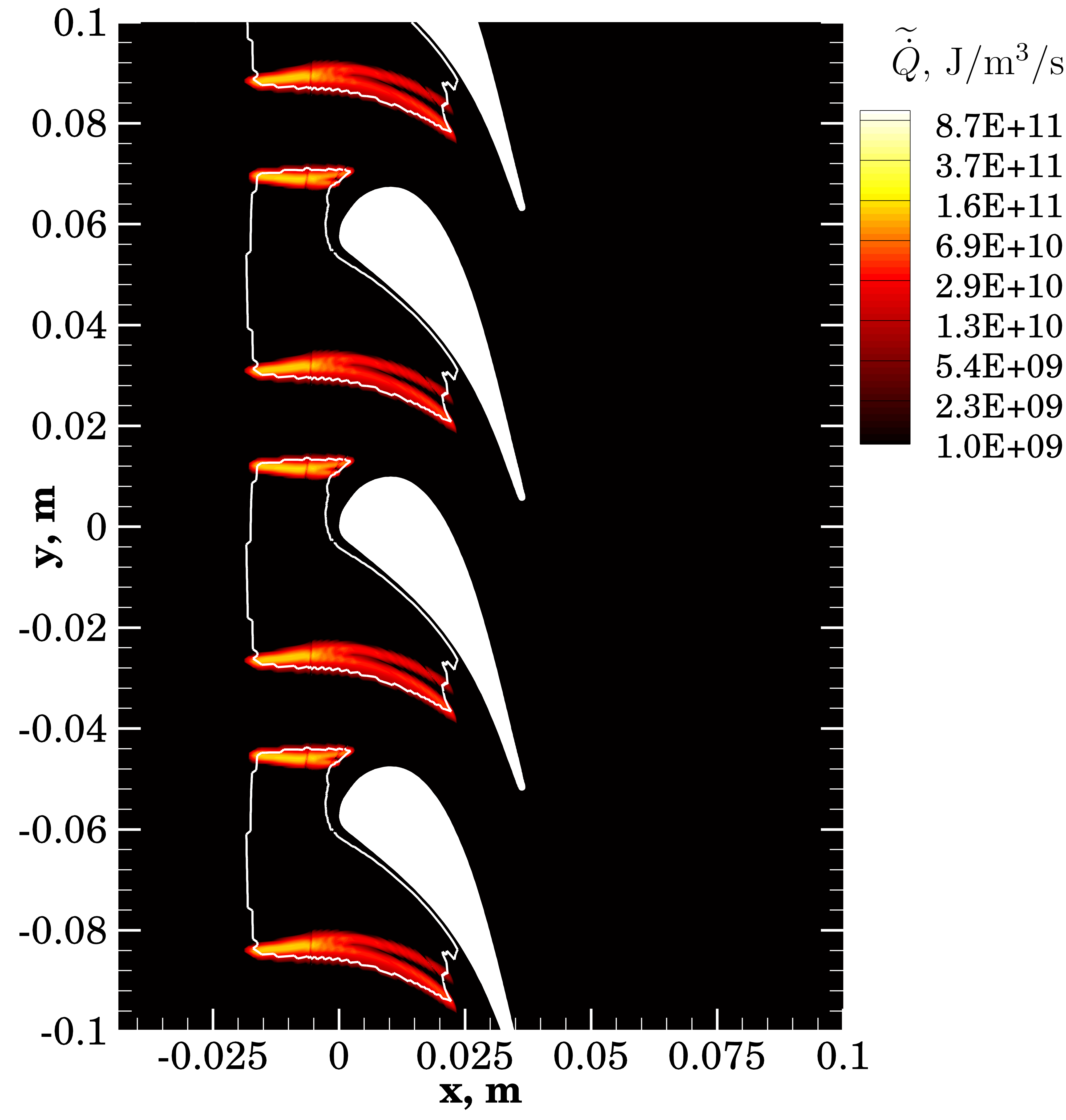}
  \caption{$\boldsymbol{\epsilon}$-based flamelet model.}
  \label{fig:q_ch4_flamelet}
\end{subfigure}%
\caption{$\boldsymbol{\widetilde{\dot{Q}}}$ fields for the $\boldsymbol{{\mathrm{CH_4}}}$ configuration.}
\label{fig:q_ch4}
\end{figure}

A deeper understanding of the combustion process can be obtained by examining the heat release rates, $\widetilde{\dot{Q}}$. Figure~\ref{fig:q_ch4} shows these quantities for both the OSK and $\epsilon$-based flamelet combustion models. In the OSK case, the reaction zones follow the flame locations shown in Fig.~\ref{fig:t_ch4_osk}, with two distinct reaction regions extending from the inlet through the passage and merging downstream of the trailing edge. The strongest reaction occurs upstream of the blade, where the pressure is highest (30~bar) and nearly uniform. As the flow approaches the blade leading edge (located at $x = 0$~cm), the presence of the blade disturbs the flow field and induces a streamwise pressure decrease. The resulting favorable pressure gradient suppresses the reaction rate along the streamwise direction. Downstream of the trailing edge, in the absence of wall constraints, the reacting gases from both flames interact with the low-speed wakes originating from the suction and pressure surfaces. This region promotes enhanced molecular and turbulent diffusion, leading to a secondary increase in reaction rates as the two flames converge. Further downstream, reactant dilution, cooling and reduced pressure cause reaction rates to drop drastically by several orders of magnitude, suppressing the heat-release–temperature feedback. As a result, effective quenching occurs despite incomplete reactant consumption. This behavior is evidenced by the abrupt decrease in $\widetilde{\dot{Q}}$ at approximately $x = 0.05$~m. Downstream of this location, although $\widetilde{\dot{Q}}$ is not strictly zero, the associated energy release is insufficient to sustain a stable flame.

In contrast, the flamelet model results exhibit a substantially different behavior. The reaction is confined to a narrow region bounded by the white contour lines in Fig.~\ref{fig:q_ch4_flamelet}. These contours mark the flamelet flammability limit, that is, the maximum attainable $S^*$ before quenching, which depends on pressure (see Figs.~\ref{fig:sscurves} and \ref{fig:tables}). Regions enclosed by these contours correspond to local values of $S^*$, as determined from $\epsilon$ via Eq.~(\ref{eq:sstar-eps}), that are below the flammability limit, allowing the flamelet to exist in a burning state. Outside these contours, $S^*$ exceeds the flammability limit and the flamelet is quenched. This flamelet quenching behavior leads directly to the distinct flame stand-off and the significantly weaker flames downstream of the blade leading edge when compared with the OSK results.

\begin{figure}
    \centering
    \includegraphics[width=0.4\linewidth]{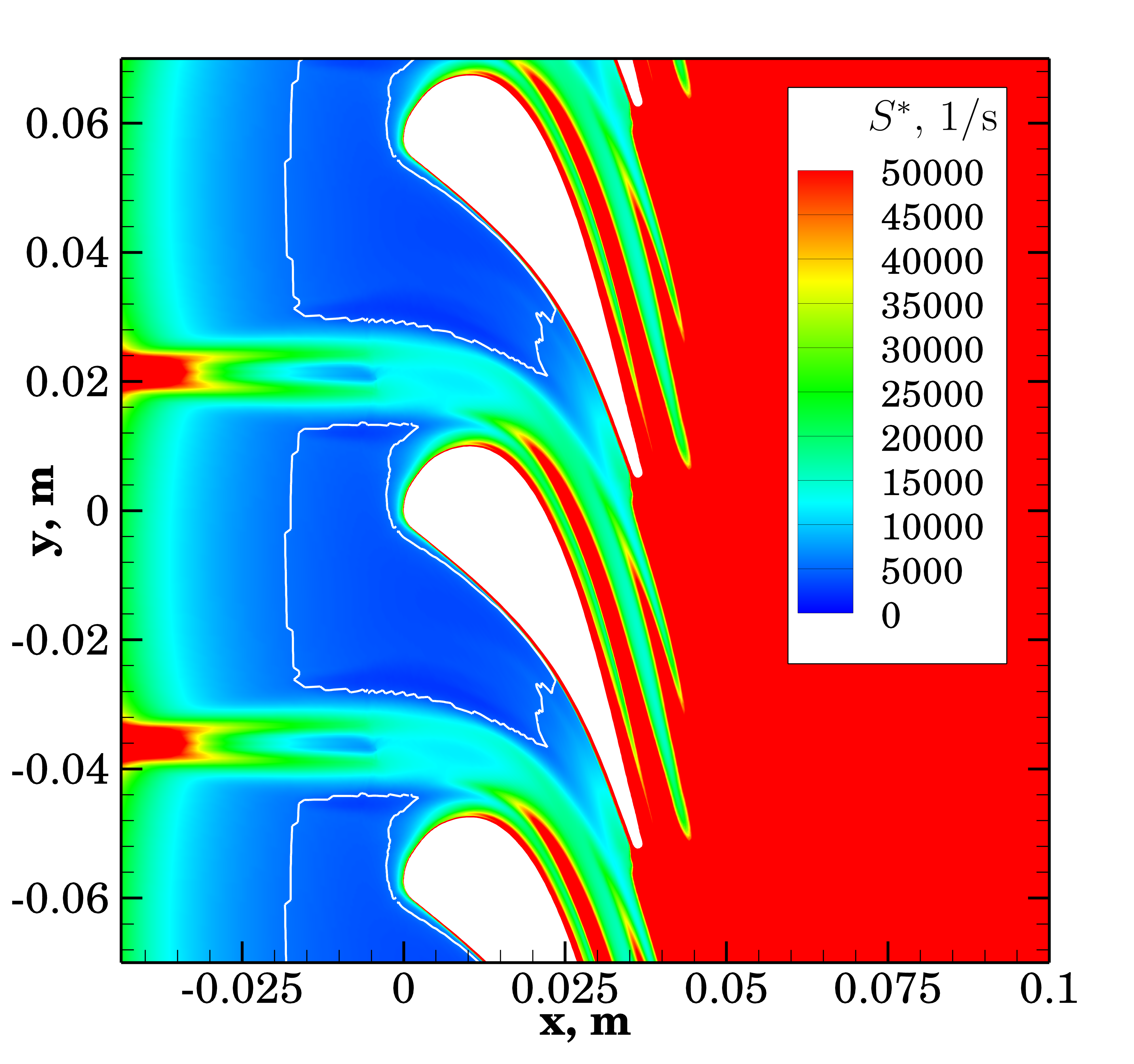}
    \caption{Flamelet inlet strain rate, $\mathbf{S^*}$, for the $\boldsymbol{\mathrm{CH_4}}$ flamelet results.}
    \label{fig:ch4_sstar}
\end{figure}

Figure~\ref{fig:ch4_sstar} shows the subgrid flamelet inflow strain-rate field, $S^*$, as determined from $\epsilon$. The same flammability-limit contours shown in Fig.~\ref{fig:q_ch4_flamelet} are overlaid here. Moderate $S^*$ values appear along the inlet (green region in Fig.~\ref{fig:ch4_sstar}) due to the specified freestream turbulence intensity, defined by the inlet values of $k$ and $\omega$. A region of high strain rate (red region) forms at the fuel-jet interface, caused by the strong velocity gradients between the two reactant streams. Downstream of the inlet, the freestream $S^*$ decreases in regions without shear production (blue regions in Fig.~\ref{fig:ch4_sstar}), dropping below the flammability limit and allowing the flamelet to exist in a burning, non-quenched state. This location defines the flame stand-off distance. This behavior suggests that the freestream turbulence intensity may influence the flame stand-off location, since $S^*$ is governed by the local evolution of $k$ and $\omega$, and thus a sensitivity analysis would be warranted.

Within the blade passage, the local velocity gradients induced by surface curvature enhance both molecular and turbulent diffusion, increasing the local strain rate $S^*$. As $S^*$ surpasses the flammability limit, the flamelet quenches within the passage. The stronger acceleration along the suction side leads to earlier quenching in that region, whereas the reaction zone along the pressure side persists farther downstream, as clearly shown in Fig.~\ref{fig:ch4_sstar}. Elevated strain rates within the near-wall boundary layers significantly exceed the flammability limit, preventing the existence of reacting flamelets close to the blade surface, which is relevant from a thermal loading perspective. Downstream of the trailing edge, the elevated turbulence levels in the wake yield $S^*$ values that are too high for the flamelets to reignite. As in the OSK case, the strongest reactions occur farthest upstream, where the pressure is highest, consistent with the expected decrease of reaction rates with pressure along the streamwise direction.

\begin{figure}
\centering
\begin{subfigure}{.45\textwidth}
  \centering
  \includegraphics[width=1.0\textwidth]{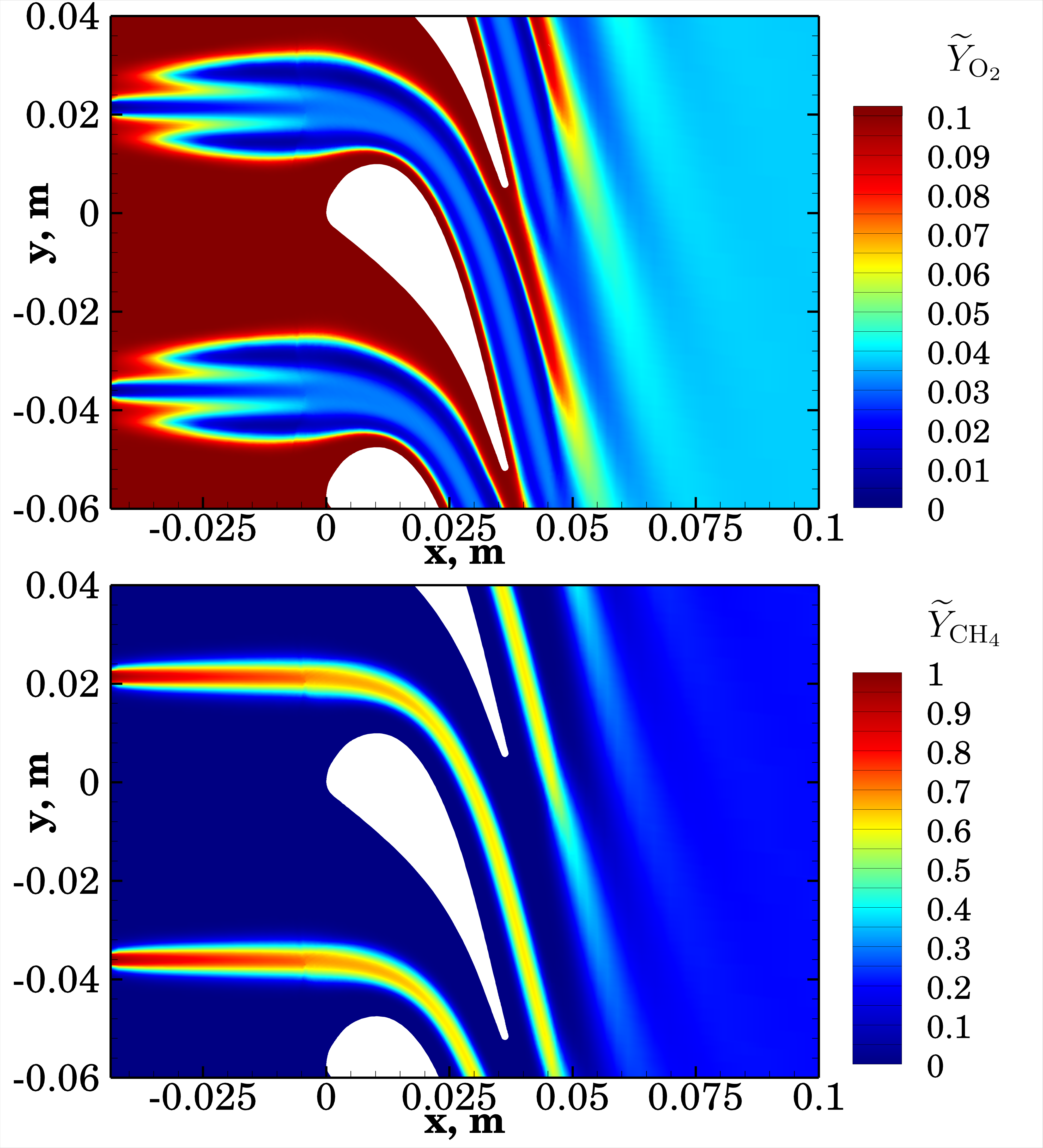}
  \caption{OSK.}
  \label{fig:osk_o2}
\end{subfigure}
\begin{subfigure}{.45\textwidth}
  \centering
  \includegraphics[width=1.0\textwidth]{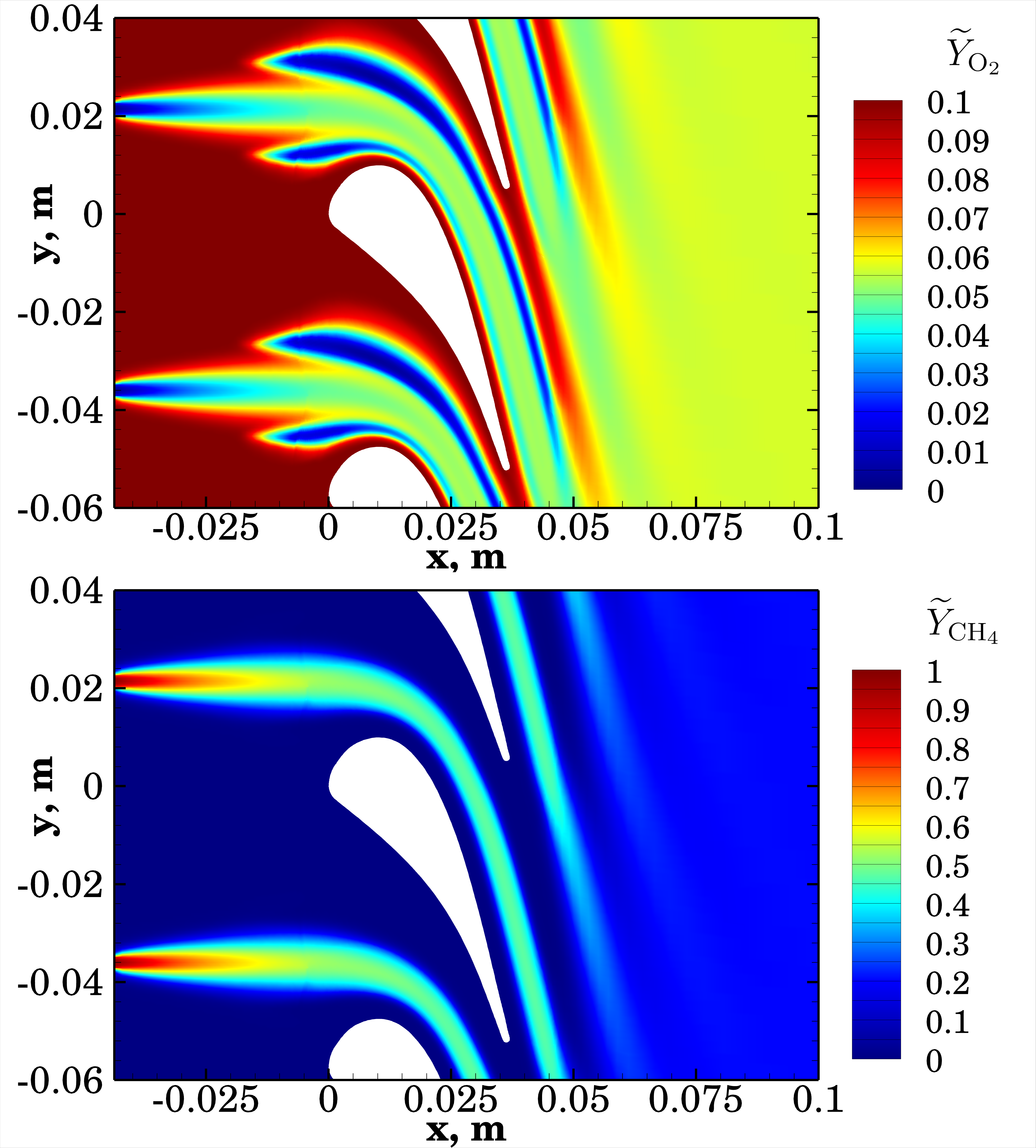}
  \caption{$\boldsymbol{\epsilon}$-based flamelet model.}
  \label{fig:ch4_o2_flamelet}
\end{subfigure}%
\caption{$\boldsymbol{\widetilde{Y}_{\mathrm{O_2}}}$ and $\boldsymbol{\widetilde{Y}_{\mathrm{CH_4}}}$ fields for the $\boldsymbol{{\mathrm{CH_4}}}$ configuration.}
\label{fig:o2_ch4}
\end{figure}

This flamelet strain rate based quenching behavior, occurring both near the inlet and within the passage, causes the $\epsilon$-based model to exhibit an overall weaker reaction. In Fig.~\ref{fig:ch4_t_fields}, this is reflected by a lower outlet temperature, approximately 100~K below that predicted by the OSK model. Figure~\ref{fig:o2_ch4} presents the $\widetilde{Y}_{\mathrm{O_2}}$ and $\widetilde{Y}_{\mathrm{CH_4}}$ fields for both combustion models. The reduced $\mathrm{O_2}$ consumption in the flamelet model further indicates weaker combustion, resulting in a larger amount of oxidizer at the outlet compared with the OSK case. In the OSK approach, downstream effective quenching occurs due to suppression of the heat-release–temperature feedback, while in the flamelet model quenching is governed by strain-rate limits. In both cases, incomplete combustion leads to residual $\mathrm{CH_4}$ at the outlet, which is undesirable from a turbine–burner perspective and suggests that fuel injection should be tuned to minimize unburned fuel. 

At the same time, the increased availability of $\mathrm{O_2}$ at the outlet in the flamelet case is relevant for multi-stage turbine–burner designs. Liu and Sirignano~\cite{liu_turbojet_2001} showed that turbine–burner performance can be enhanced using a multi-stage configuration, in which additional fuel is injected in downstream stages to sustain combustion across successive stages. Such configurations require sufficient oxidizer to remain available beyond the first stage. Consequently, the differences observed between the OSK and $\epsilon$-based flamelet models are of direct practical relevance and may lead to substantially different performance once the rotor and multiple staged are considered.

\subsection{JP-5 Configuration}

\begin{figure}
\centering
\begin{subfigure}{.45\textwidth}
  \centering
  \includegraphics[width=1.0\textwidth]{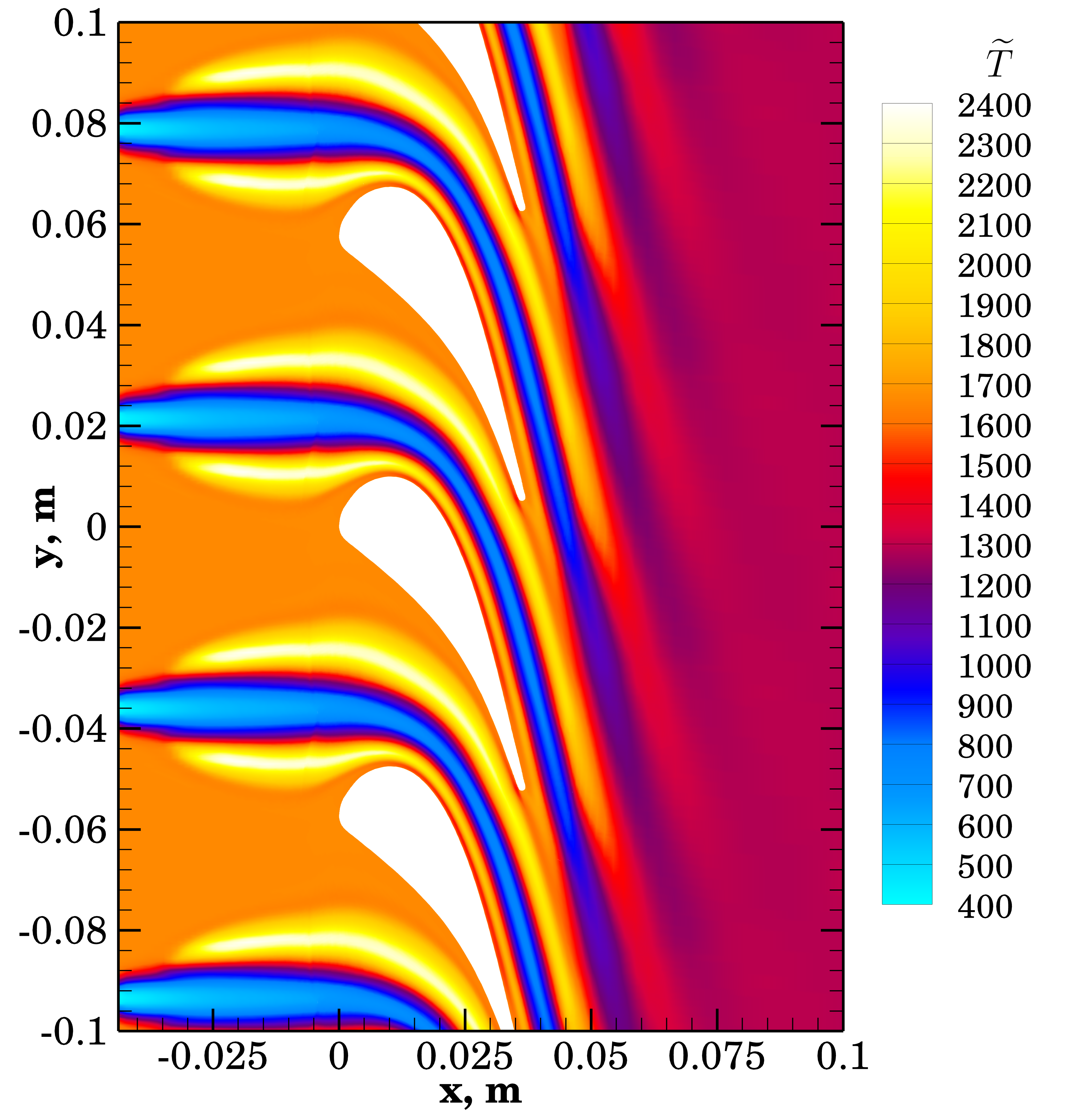}
  \caption{$\mathbf{\widetilde{T}}$.}
  \label{fig:t_jp5}
\end{subfigure}
\begin{subfigure}{.45\textwidth}
  \centering
  \includegraphics[width=0.99\textwidth]{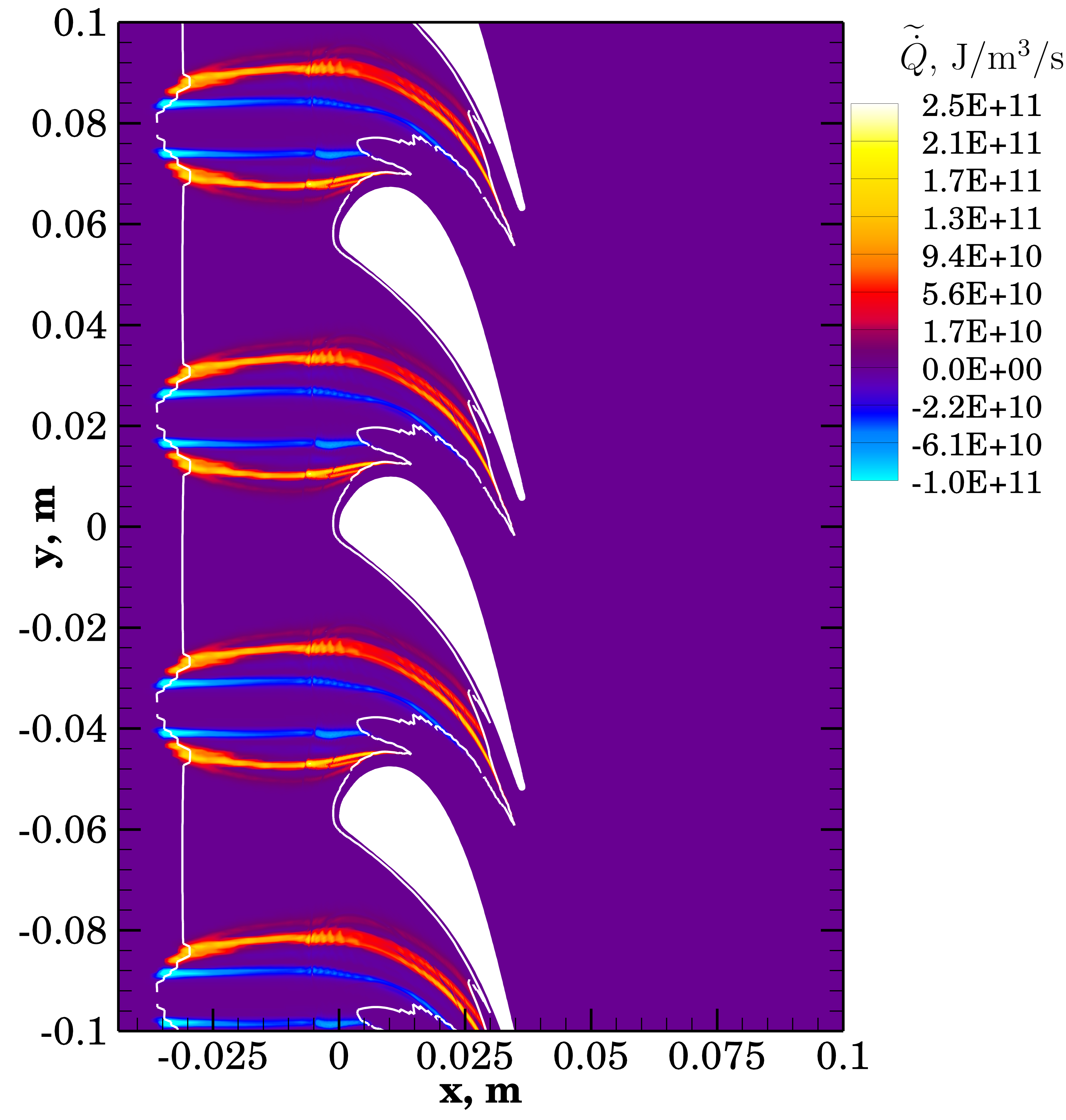}
  \caption{$\boldsymbol{\widetilde{\dot{Q}}}$.}
  \label{fig:q_jp5}
\end{subfigure}%
\caption{$\mathbf{\widetilde{T}}$ and $\boldsymbol{\widetilde{\dot{Q}}}$ fields for the JP-5 configuration.}
\label{fig:jp5_t_and_q}
\end{figure}

\begin{figure}
    \centering
    \includegraphics[width=1.0\linewidth]{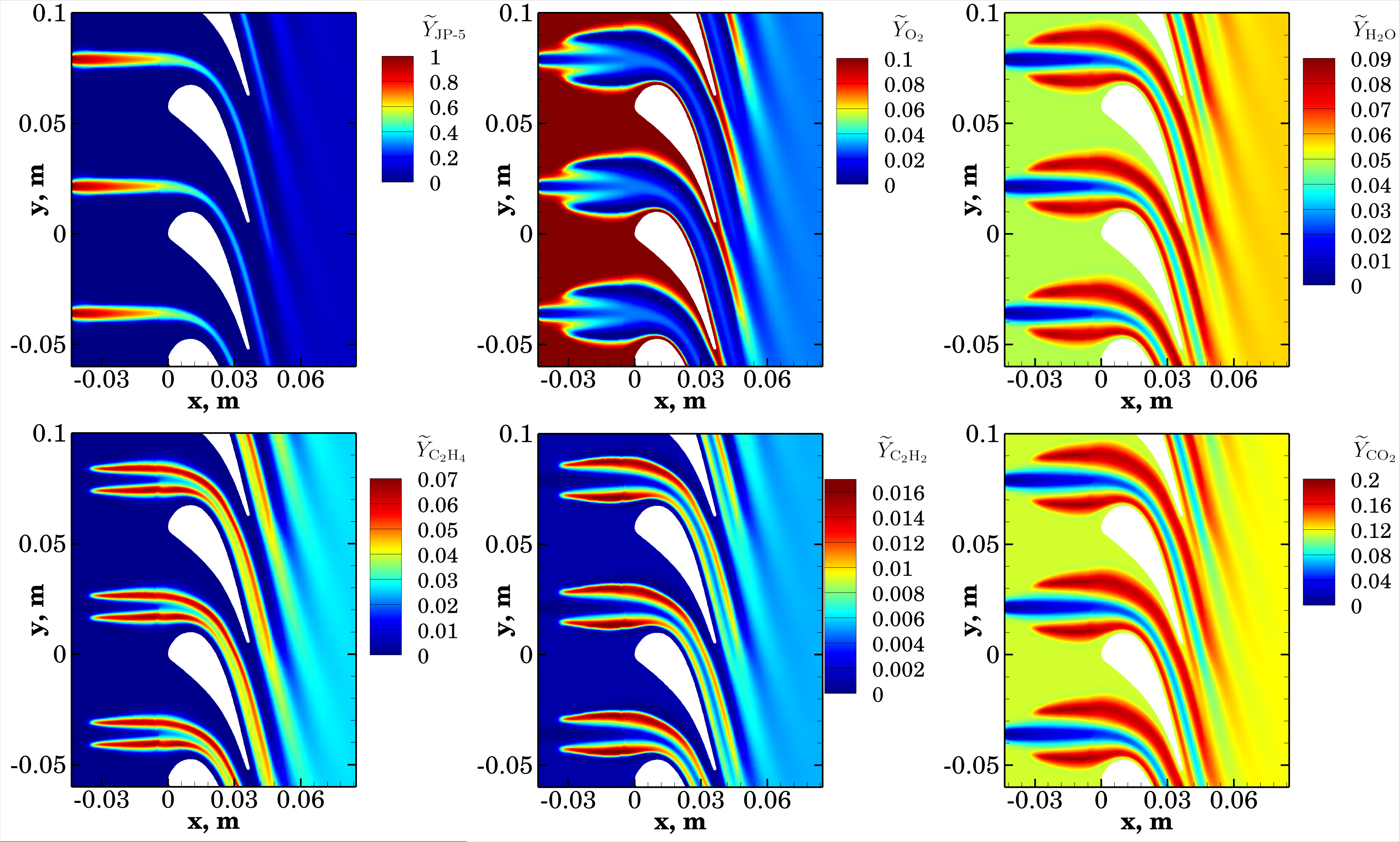}
    \caption{Mixture $\boldsymbol{\widetilde{Y}_n}$ for the JP-5 configuration. From top to bottom and from left to right: JP-5, $\boldsymbol{\mathrm{O_2}}$, $\boldsymbol{\mathrm{H_2O}}$, $\boldsymbol{\mathrm{C_2H_4}}$, $\boldsymbol{\mathrm{C_2H_2}}$ and $\boldsymbol{\mathrm{CO_2}}$.}
    \label{fig:jp5_compositions}
\end{figure}

 Figure \ref{fig:t_jp5} reports the mean temperature contours for the JP-5 configuration. The combustion behavior of JP-5 is qualitatively similar to that of the $\mathrm{CH_4}$ $\epsilon$-based results. The same two-flame structure, consisting of suction-side and pressure-side flames, appears. As in the $\mathrm{CH_4}$ flamelet results, the JP-5 flames weaken within the passage and remain disjointed downstream of the trailing edge. For comparable mass-flow conditions, JP-5 combustion attains higher peak flame temperatures, reaching approximately 2400~K, compared with 2200–2300~K for the $\mathrm{CH_4}$ configurations. (see Fig.~\ref{fig:ch4_t_fields}). The outlet temperature downstream of the passage is also higher by roughly 100–200~K. Similar to the $\mathrm{CH_4}$ flamelet case, the JP-5 configuration exhibits a distinct flame stand-off, though less pronounced, with ignition occurring approximately 1~cm downstream of the inlet. Again, this stand-off promotes additional mixing of the reactants before ignition, leading to wider downstream reaction zones when compared to the OSK model.

Figure~\ref{fig:q_jp5} shows the $\widetilde{\dot{Q}}$ field for the JP-5 configuration. The same flammability-limit contour lines used in Figs.~\ref{fig:q_ch4_flamelet} and \ref{fig:ch4_sstar} are applied here. Notably, both endothermic and exothermic reactions are transferred from the flamelet to the resolved scale within the flammability region. The endothermic zones (negative $\widetilde{\dot{Q}}$, indicated by the blue regions) correspond to the pyrolysis of JP-5, which produces lighter hydrocarbons. The red regions represent the exothermic zones (positive $\widetilde{\dot{Q}}$) associated with the subsequent oxidation of these hydrocarbons.

A key difference relative to the $\mathrm{CH_4}$ case lies in the location of the flammability limits. Due to the higher flammability limit of JP-5 (see Figs.~\ref{fig:sscurves} and \ref{fig:q_jp5_tables}), combustion occurs earlier, resulting in a shorter flame stand-off distance. For the same reason, the reaction extends farther into the passage compared with the $\mathrm{CH_4}$ configuration. The pressure-side reaction zone also penetrates farther downstream than the suction-side zone, owing to the lower flow acceleration and consequently lower strain rates on the pressure side. These results demonstrate that the resolved-scale quenching behavior is consistent with the subgrid-scale flamelet flammability limits.

Figure~\ref{fig:jp5_compositions} shows the resolved-scale mass-fraction fields for several representative species. Similar to the subgrid behavior reported in Fig.~\ref{fig:yn_tab}, JP-5 originating from the fuel stream undergoes pyrolysis, decomposing into lighter hydrocarbons (here $\mathrm{C_2H_4}$ and $\mathrm{C_2H_2}$ are shown) on both sides of the JP-5 jet. Once these intermediate hydrocarbons are formed, they react with $\mathrm{O_2}$ from the vitiated-air stream, releasing heat through exothermic oxidation and producing combustion products.

The $\mathrm{H_2O}$ and $\mathrm{CO_2}$ fields reveal a clear vertical displacement of the product zones relative to the JP-5 stream. This occurs because the JP-5 must first decompose into smaller hydrocarbons before oxidation can proceed, creating a region between the unreacted JP-5 and the final products where intermediates such as $\mathrm{C_2H_4}$ and $\mathrm{C_2H_2}$ reside. As a result, the exothermic reaction zones are shifted closer to the blade surfaces on both the suction and pressure sides compared with the $\mathrm{CH_4}$ configuration.

\subsection{Global Performance Metrics}

\begin{figure}[h]
\centering
\begin{subfigure}{.45\textwidth}
  \centering
  \includegraphics[width=0.97\textwidth]{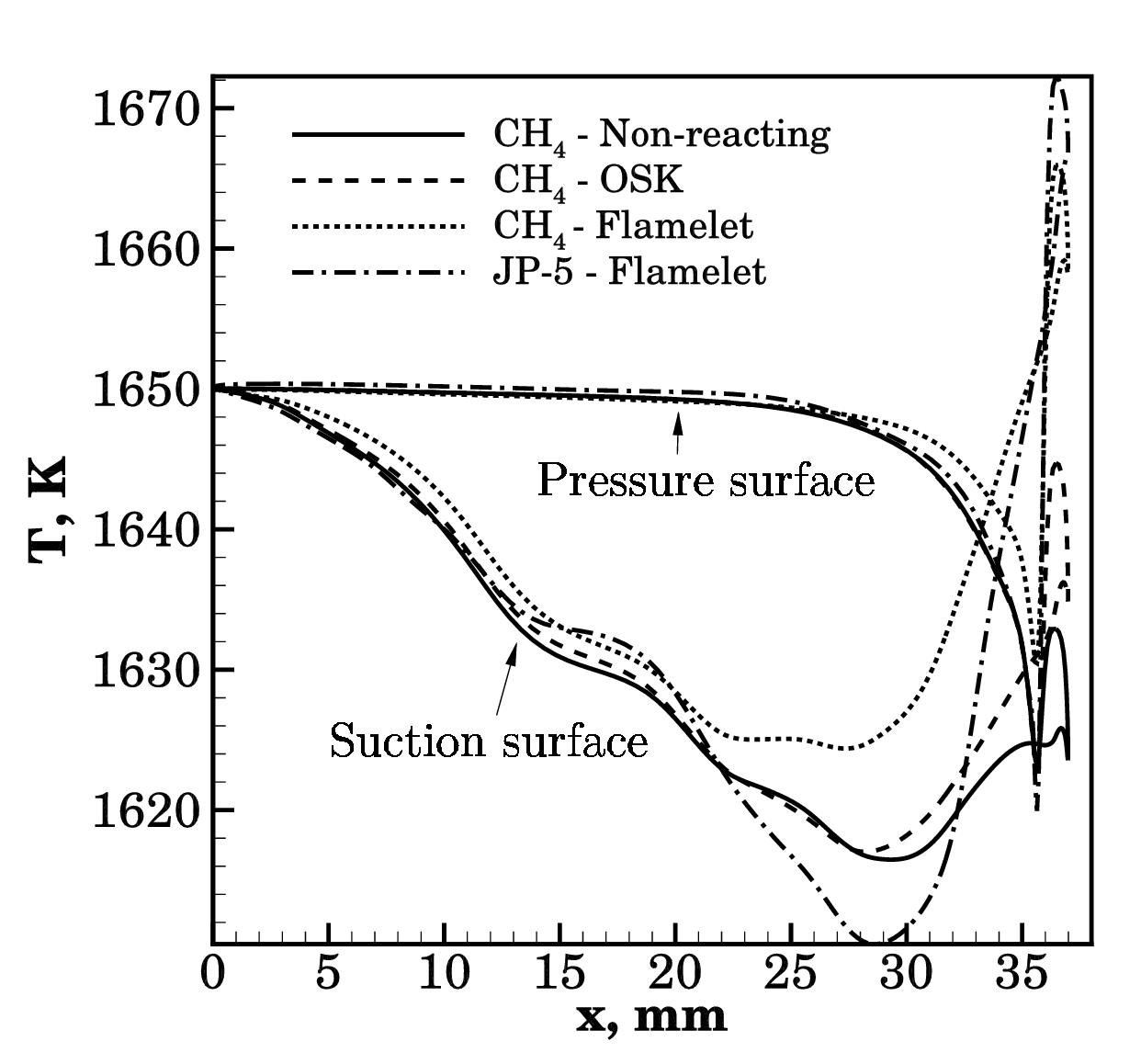}
  \caption{Blade surface temperatures.}
  \label{fig:wall_temp}
\end{subfigure}
\begin{subfigure}{.45\textwidth}
  \centering
  \includegraphics[width=1.0\textwidth]{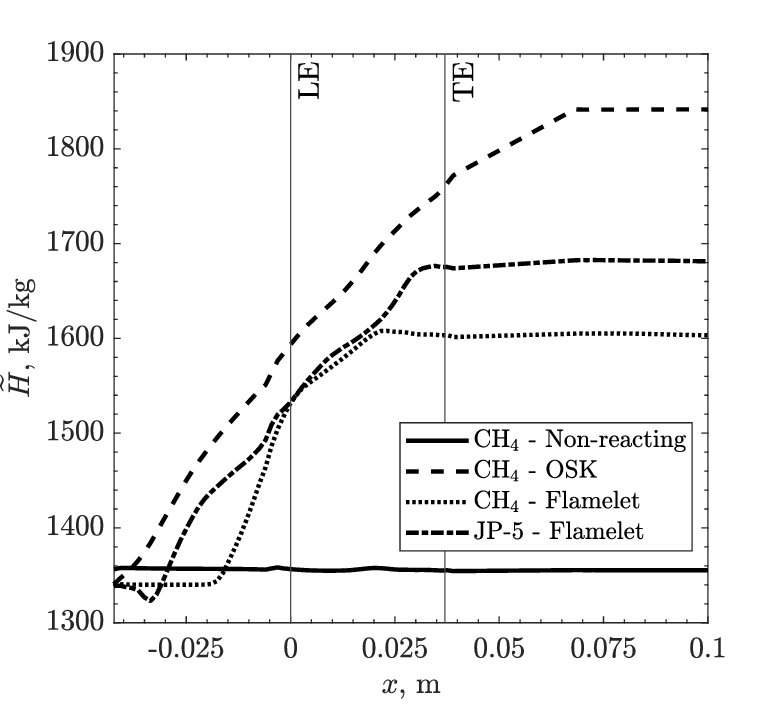}
  \caption{Axial surface-averaged total sensible enthalpy, $\mathbf{\widetilde{H}}$.}
  \label{fig:H}
\end{subfigure}%
\caption{Blade surface temperature distribution and axial variations of surface-averaged total sensible enthalpy, $\mathbf{\widetilde{H}}$. }
\label{fig:wall_and_H}
\end{figure}

Figure~\ref{fig:wall_temp} shows the static temperature along the suction and pressure surfaces for all reacting cases, together with a non-reacting $\mathrm{CH_4}$ baseline where chemical rates are zero. Although the walls are adiabatic and no heat-transfer model is included, the results illustrate relative near-wall temperature trends. In the non-reacting case, wall temperature decreases along the chord from the inlet vitiated-air value of 1650 K due to flow acceleration through the passage. All reacting cases exhibit a similar initial decrease, followed by temperature recovery near the trailing edge where the flames approach the blade. Despite the stronger reaction compared to the flamelet configurations, the OSK model yields a trailing-edge wall temperature only about 10 K above the non-reacting case. In contrast, the $\epsilon$-based flamelet model produces higher trailing-edge temperatures, reaching approximately 1670 K, or about 40 K above the non-reacting case, due to the broader reaction zone associated with flame stand-off. The JP-5 case exhibits the highest wall temperatures, reflecting both the wider reaction zone and the vertical displacement of combustion induced by fuel pyrolysis which brings the flames closer to the blade.

Figure \ref{fig:H} shows the axial variations of surface-averaged total sensible enthalpy, $\widetilde{H}=\widetilde{E}+\bar{p}/\bar{\rho}$, which excludes enthalpy of formation. The quantities are computed over streamwise cross-sections of the computational domain and are shown for all reacting cases, as well as for the non-reacting $\mathrm{CH_4}$ case. The vertical lines indicate the blade leading and trailing edge locations (LE and TE). In the non-reacting case, the total sensible enthalpy remains conserved, as expected in the absence of chemical heat release. Small differences in inlet enthalpy between configurations arise from slight variations in the mass flow rate, $\dot{m}$, reported in Table~\ref{tab:energy_balance}, which result from density differences associated with the differing downstream temperatures across cases.

For the OSK configuration, the enthalpy increases immediately, consistent with the absence of flame stand-off. The enthalpy continues to rise through the blade passage and reaches a plateau shortly downstream of the trailing edge, where reaction rates decrease substantially. In contrast, the $\epsilon$-based flamelet cases exhibit a delayed enthalpy increase corresponding to the flame stand-off. In these cases, $\widetilde{H}$ plateaus within the passage as combustion quenches due to strain-rate-induced flamelet extinction. Notably, the JP-5 case exhibits an initial decrease in enthalpy, which is attributed to endothermic fuel pyrolysis prior to oxidation.

For the steady, time-averaged flow with adiabatic walls, integration of the resolved-scale energy equation (Eq. (\ref{eq:energy})) over the computational domain yields the balance
\begin{equation}
    \dot{m}\widetilde{H}|_{\mathrm{out}}-\dot{m}\widetilde{H}|_{\mathrm{in}} + W_f= Q \:,
\end{equation}
where $Q = \int_\Omega\widetilde{\dot{Q}}d\Omega$ represents the total heat released by chemical reaction and $W_f$ represents the residual losses associated with viscous and wall effects, which is inferred to be small from the near-zero enthalpy change in the non-reacting case. Normalizing by the mass flow rate gives
\begin{equation}
     \Delta\widetilde{H}=\widetilde{H}|_{\mathrm{out}}-\widetilde{H}|_{\mathrm{in}} =(Q-W_f)/\dot{m} \:,
\end{equation}
representing the net energy added per unit mass by combustion (and small viscous loss contribution). This quantity directly represents the energy available for extraction by the downstream rotor and is summarized for all cases in Table~\ref{tab:energy_balance}.

\setcounter{table}{0}
\begin{table}
\caption{\label{tab:energy_balance} Energy balance in the turbine stator passage.}
\centering
\begin{tabular}{lcccccc}
\hline
  & $\dot{m}$, kg/m/s & $\widetilde{H}|_{\mathrm{in}}$, kJ/kg&$\widetilde{H}|_{\mathrm{out}}$, kJ/kg& $\Delta\widetilde{H}$, kJ/kg& \%$\Delta \widetilde{H}=100\frac{\widetilde{H}|_{\mathrm{out}}-\widetilde{H}|_{\mathrm{out,NR}}}{\widetilde{H}|_{\mathrm{out,NR}}}$\\\hline
$\mathrm{CH_4}$ - Non-reacting  & 46.0& 1355.6& 1355.4& -0.2 & -\\
$\mathrm{CH_4}$ - OSK & 43.0& 1334.1& 1841.6& 507.5 & 35.9\\
$\mathrm{CH_4}$ - Flamelet& 44.4& 1340.8& 1603.3& 262.5& 18.3\\
JP-5 - Flamelet& 47.0& 1335.2& 1681.2& 346.0&24.0\\
\hline
\end{tabular}
\end{table}
The last column in Table~\ref{tab:energy_balance} reports the percentage increase in added energy due to combustion relative to the non-reacting case. For the non-reacting configuration, $Q=0$, and the small difference between inlet and outlet enthalpy therefore reflects numerical dissipation and viscous losses, represented by $W_f$. For the same fuel ($\mathrm{CH_4}$), the $\epsilon$-based flamelet model predicts approximately half of the added energy obtained with the OSK model, indicating a substantially weaker overall reaction. This reduction is consistent with the smaller reacting regions imposed by strain-rate-based flamelet quenching. Despite capturing endothermic pyrolysis effects, the JP-5 configuration yields a larger net energy addition than the $\mathrm{CH_4}$ $\epsilon$-based case. This result is noteworthy because JP-5 has a lower specific heat of combustion (approximately 42–43 MJ/kg) than $\mathrm{CH_4}$ (approximately 50 MJ/kg). The increased energy addition arises from the higher strain-rate flammability limit of the JP-5 flamelets, which permits larger reacting regions on the resolved scale, manifested as shorter flame stand-off distances and delayed quenching within the passage relative to $\mathrm{CH_4}$. Nevertheless, the net energy addition for JP-5 remains lower than that predicted by the OSK $\mathrm{CH_4}$ model, reaching approximately two-thirds of the OSK value.

\begin{figure}
    \centering
    \includegraphics[width=0.8\linewidth]{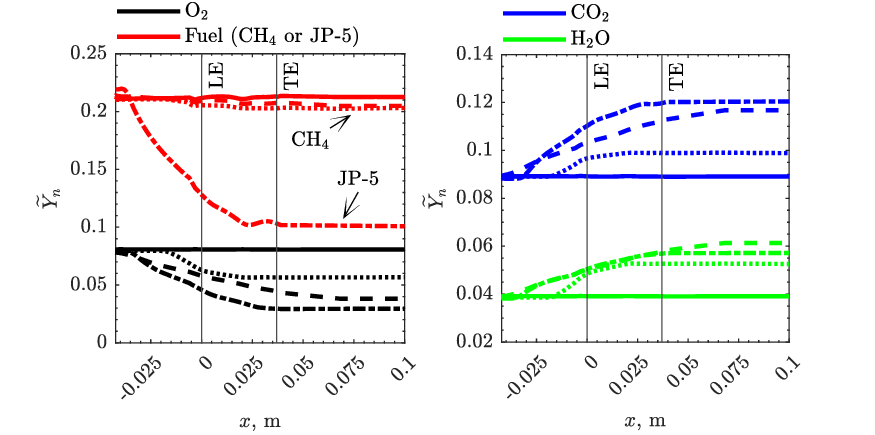}
    \caption{Axial variations of surface-averaged reactant and product mass fractions. Solid lines: $\boldsymbol{\mathrm{CH_4}}$ - non-reacting configuration; dashed lines: $\boldsymbol{\mathrm{CH_4}}$ - OSK configuration; dotted lines $\boldsymbol{\mathrm{CH_4}}$ - flamelet configuration; dash dot lines: JP-5 - flamelet configuration}
    \label{fig:y_averages}
\end{figure}

Figure~\ref{fig:y_averages} shows axial variations of surface-averaged reactant and product mass fractions for the non-reacting $\mathrm{CH_4}$ case and all reacting configurations. Vertical lines denote the blade leading and trailing edges. In the non-reacting case, mass fractions remain constant, with small deviations near the leading edge attributed to interpolation artifacts.
In reacting cases, reactant mass fractions decrease and product mass fractions increase until plateauing downstream of the reaction zone. These transition locations coincide with changes in $\widetilde{H}$ and reflect differences in flame stand-off and quenching behavior. Overall combustion is weak due to the low $\mathrm{O_2}$ content of the vitiated air, with major product mass fractions increasing by only about 2-3\%. In the $\mathrm{CH_4}$ cases, the fuel mass fraction decreases by approximately 1\%, leaving a substantial amount of excess fuel. Fuel injection may be tuned to avoid this undesirable situation. While both combustion models consume similar amounts of $\mathrm{CH_4}$, the $\epsilon$-based flamelet model produces less $\mathrm{CO_2}$, as a portion of the fuel carbon is distributed among intermediate species such as $\mathrm{CO}$, $\mathrm{CH_3}$, and $\mathrm{CH_2O}$.

In contrast, the JP-5 mass fraction decreases by approximately 50\% (from about 0.2 to 0.1) as the fuel undergoes pyrolysis into lighter hydrocarbons. A modeling artifact associated with the $\epsilon$-based flamelet approach is observed downstream of the trailing edge. In this region, temperatures remain near 1300 K, while JP-5 pyrolysis at this pressure occurs at temperatures above approximately 900 K. Physically, no residual JP-5 should therefore persist downstream. However, due to upstream strain-rate-induced quenching, the flamelet model returns zero chemical source terms downstream, preventing further endothermic pyrolysis and leading to an overprediction of remaining JP-5 mass fraction and underprediction of the generated lighter hydrocarbons.

Walsh et al.~\cite{walsh_performance_2026} applied the same $\epsilon$-based flamelet model to $\mathrm{CH_4}$ combustion with oxygen-enriched air (50\% $\mathrm{O_2}$ and 50\% $\mathrm{N_2}$ by mass) in an accelerating two-dimensional mixing layer subject to a purely axial favorable pressure gradient of comparable magnitude. In the present study, the use of vitiated air with reduced $\mathrm{O_2}$ concentration leads, beyond the expected reductions in reaction intensity and peak temperature, to a decrease of approximately two orders of magnitude in the flamelet strain-rate flammability limit. This results in smaller reacting regions in the direction of the pressure gradient due to earlier quenching. Moreover, the presence of both axial and transverse pressure gradients in the turbine passage produces a more complex $S^*$ field, with quenching locations varying in both the streamwise and transverse directions.

\section{Conclusions}
Combustion within a two-dimensional turbine stator passage has been numerically simulated using a RANS framework coupled with the novel $\epsilon$-based flamelet model. This model links resolved-scale turbulence quantities with subgrid flamelet dynamics through the local turbulent kinetic energy dissipation rate, $\epsilon$, which determines the flamelet inflow strain rate, $S^*$.

The $\epsilon$-based flamelet model has, for the first time, enabled the use of the practical JP-5 as fuel within the turbine-burner design concept. This was achieved by solving transport equations for 14 species on the resolved scale, while chemical source terms are obtained from a precomputed flamelet library incorporating finite-rate kinetics modeled with the HyChem A3 mechanism, which includes 119 species and 841 elementary reactions.

The performance of the $\epsilon$-based model was evaluated through a comparison of $\mathrm{CH_4}$–vitiated air combustion using a one-step kinetic (OSK) combustion model and an $\epsilon$-based approach employing a 13-species skeletal reduction of the FFCM-1 mechanism. Relative to the OSK model, the $\epsilon$-based flamelet formulation predicts lower peak flame temperatures, by approximately 100 K, due to the inclusion of dissociation effects. Flame stand-off distances naturally arise from the strain-rate-based quenching constraint, promoting enhanced mixing but weaker overall reaction intensity. The predicted quenching locations correlate with regions of high strain and flow acceleration within the blade passage. Comparisons were made with results for an accelerating-mixing-layer study where the air was not vitiated. Substantial reduction in flammability limits occurred for methane combustion with vitiated air. This requires special attention for the accelerating flow through the stator passage.

For the JP-5 configuration, the model captures both endothermic pyrolysis and exothermic oxidation processes, resulting in vertically displaced reaction zones and higher near-wall temperatures relative to the $\mathrm{CH_4}$ cases. Compared with the $\mathrm{CH_4}$ flamelet configuration, JP-5 exhibits shorter flame stand-off distances and delayed quenching within the passage, owing to its higher strain-rate-based flammability limit, which permits larger reacting regions prior to quenching.

Global performance metrics further highlight these differences. For the same fuel, the $\epsilon$-based flamelet model yields approximately 50\% lower net energy addition and increased oxidizer availability at the outlet compared with the OSK model, due to early strain-rate-based quenching, with implications for multi-stage turbine-burner operation. Near-wall temperature trends indicate higher trailing-edge temperatures for the flamelet cases than for the OSK model, associated with broader reaction zones resulting from flame stand-off, while remaining below OSK-predicted peak levels. When comparing $\epsilon$-based flamelet models for $\mathrm{CH_4}$ and JP-5, the JP-5 case produces greater net energy addition than the $\mathrm{CH_4}$ flamelet configuration, despite its lower specific heat of combustion, due to its larger resolved-scale reaction regions.

\section*{Acknowledgments}
The research was supported by the Office of Naval Research through Grant N00014-22-1-2467 with Dr. Steven Martens as program manager. Professor Heinz Pitsch of RWTH Aachen University is acknowledged for providing us access to the FlameMaster code. Artificial intelligence tools were used solely to improve grammar and language clarity during manuscript preparation. All technical content, figures, and analyses are original.

\bibliography{references,referances_2}

@misc{flamemaster,
   author = {H. Pitsch},
   title = {FlameMaster, A C++ Computer Program for 0D Combustion and 1D Laminar Flame Calculations. Version V4.2.1},
   url = {https://www.itv.rwth-aachen.de/downloads/flamemaster/},
   year = {2022},
}

@article{zhu_simulation_2025,
	title = {Simulation of {Multi}-{Injector} {H2}-{O2} {Rocket} {Combustion} {Instability}},
	volume = {Article in Advance},
	doi = {10.2514/1.J065542},
	journal = {AIAA Journal},
	author = {Zhu, Yalu and Liu, Feng and Sirignano, William A},
	year = {2025},
}

@incollection{menter_ten_2003,
	address = {Danbury, CT},
	title = {Ten {Years} of {Industrial} {Experience} with the {SST} {Turbulence} {Model}},
	booktitle = {Turbulence, {Heat} and {Mass} {Transfer} 4},
	publisher = {Begell House},
	author = {Menter, Florian R and Kuntz, Martin and Langtry, Robin},
	editor = {Hanjalic, K and Nagano, Y and Tummers, M},
	year = {2003},
	pages = {625--632},
}

@inproceedings{zhu_large-eddy_2025,
	address = {Orlando, Florida},
	title = {Large-{Eddy} {Simulation} of {Reacting} {Flow} in a {Turbine} {Stage}},
	doi = {10.2514/6.2025-1129},
	booktitle = {{AIAA} {SCITECH} 2025},
	author = {Zhu, Yalu and Walsh, Sylvain L and Liu, Feng and Sirignano, William A},
	month = jan,
	year = {2025},
	note = {Paper 2025-1129},
}

@inproceedings{walsh_performance_2026,
	address = {Orlando, Florida},
	title = {Performance of {Flamelet} {Models} with {Epsilon} {Tracking} for {Diffusion} {Flame} {Simulations}},
	booktitle = {{AIAA} {SCITECH} 2026 {Forum}},
	author = {Walsh, Sylvain L. and Zhu, Yalu and Liu, Feng and Sirignano, William A.},
	year = {2026},
}

@article{zhu_numerical_2024,
	title = {Numerical {Investigation} of {Diffusion} {Flame} in {Transonic} {Flow} with {Large} {Pressure} {Gradient}},
	volume = {40},
	issn = {0748-4658},
	doi = {10.2514/1.B39341},
	number = {4},
	urldate = {2024-05-22},
	journal = {Journal of Propulsion and Power},
	author = {Zhu, Yalu and Liu, Feng and Sirignano, William A.},
	month = feb,
	year = {2024},
	pages = {519--532},
}

@article{walsh_turbulent_2025,
	title = {Turbulent {Accelerating} {Combusting} {Flows} with {Methane}-{Vitiated} {Air} {Flamelet} {Model}},
	volume = {63},
	issn = {0001-1452},
	doi = {10.2514/1.J064259},
	number = {4},
	urldate = {2025-05-20},
	journal = {AIAA Journal},
	author = {Walsh, Sylvain L. and Zhan, Lei and Mehring, Carsten and Liu, Feng and Sirignano, William A.},
	month = apr,
	year = {2025},
	pages = {1474--1489},
}

@article{sirignano_diffusion_1997,
	title = {Diffusion flame in a two-dimensional, accelerating mixing layer},
	volume = {9},
	doi = {10.1063/1.869378},
	number = {9},
	journal = {Physics of Fluids},
	author = {Sirignano, William A and Kim, Inchul},
	month = sep,
	year = {1997},
	pages = {2617--2630},
}

@article{yin_review_2020,
	title = {A review of gas turbine engine with inter-stage turbine burner},
	volume = {121},
	issn = {0376-0421},
	doi = {10.1016/j.paerosci.2020.100695},
	urldate = {2024-05-22},
	journal = {Progress in Aerospace Sciences},
	author = {Yin, Feijia and Rao, Arvind Gangoli},
	month = feb,
	year = {2020},
	pages = {100695},
}

@article{peters_laminar_1988,
	title = {Laminar flamelet concepts in turbulent combustion},
	volume = {21},
	doi = {10.1016/S0082-0784(88)80355-2},
	number = {1},
	journal = {Symposium (International) on Combustion},
	author = {Peters, N.},
	month = jan,
	year = {1988},
	pages = {1231--1250},
}

@article{bilger_structure_1976,
	title = {The {Structure} of {Diffusion} {Flames}},
	volume = {13},
	issn = {0010-2202},
	doi = {10.1080/00102207608946733},
	number = {1-6},
	urldate = {2024-05-21},
	journal = {Combustion Science and Technology},
	author = {Bilger, R. W.},
	month = jul,
	year = {1976},
	pages = {155--170},
}

@article{pierce_progress-variable_2004,
	title = {Progress-variable approach for large-eddy simulation of non-premixed turbulent combustion},
	volume = {504},
	issn = {0022-1120, 1469-7645},
	doi = {10.1017/S0022112004008213},
	language = {en},
	urldate = {2024-05-16},
	journal = {Journal of Fluid Mechanics},
	author = {Pierce, Charles D. and Moin, Parviz},
	month = apr,
	year = {2004},
	pages = {73--97},
}

@phdthesis{pierce_progress_2001,
	address = {United States -- California},
	type = {Ph.{D}.},
	title = {Progress -variable approach for large -eddy simulation of turbulent combustion},
	copyright = {Database copyright ProQuest LLC; ProQuest does not claim copyright in the individual underlying works.},
	language = {English},
	urldate = {2024-05-16},
	school = {Stanford University},
	author = {Pierce, Charles David},
	year = {2001},
	note = {ISBN: 9780493383156},
}

@misc{sirignano_flamelet_2024,
	title = {Flamelet {Connection} to {Turbulence} {Kinetic} {Energy} {Dissipation} {Rate}},
	doi = {10.48550/arXiv.2409.04929},
	urldate = {2025-05-20},
	publisher = {arXiv},
	author = {Sirignano, William A. and Hellwig, Wes and Walsh, Sylvain L.},
	month = sep,
	year = {2024},
	note = {In journal review},
}

@techreport{mcbride_coefficients_1993,
	title = {Coefficients for calculating thermodynamic and transport properties of individual species},
	number = {NASA-TM-4513},
	urldate = {2025-05-21},
	author = {Mcbride, Bonnie J. and Gordon, Sanford and Reno, Martin A.},
	month = oct,
	year = {1993},
}

@book{peters_turbulent_2000,
	address = {Cambridge},
	series = {Cambridge {Monographs} on {Mechanics}},
	title = {Turbulent {Combustion}},
	isbn = {978-0-521-66082-2},
	publisher = {Cambridge University Press},
	author = {Peters, Norbert},
	year = {2000},
	doi = {10.1017/CBO9780511612701},
}

@article{janicka_two-variables_1979,
	series = {Seventeenth {Symposium} ({International}) on {Combustion}},
	title = {A two-variables formalism for the treatment of chemical reactions in turbulent {H2}—{Air} diffusion flames},
	volume = {17},
	issn = {0082-0784},
	doi = {10.1016/S0082-0784(79)80043-0},
	number = {1},
	urldate = {2024-05-21},
	journal = {Symposium (International) on Combustion},
	author = {Janicka, J. and Kollmann, W.},
	month = jan,
	year = {1979},
	pages = {421--430},
}

@article{wang_physics-based_2018,
	title = {A physics-based approach to modeling real-fuel combustion chemistry - {I}. {Evidence} from experiments, and thermodynamic, chemical kinetic and statistical considerations},
	volume = {193},
	issn = {0010-2180},
	doi = {10.1016/j.combustflame.2018.03.019},
	urldate = {2024-05-22},
	journal = {Combustion and Flame},
	author = {Wang, Hai and Xu, Rui and Wang, Kun and Bowman, Craig T. and Hanson, Ronald K. and Davidson, David F. and Brezinsky, Kenneth and Egolfopoulos, Fokion N.},
	month = jul,
	year = {2018},
	pages = {502--519},
}

@article{xu_physics-based_2018,
	title = {A physics-based approach to modeling real-fuel combustion chemistry – {II}. {Reaction} kinetic models of jet and rocket fuels},
	volume = {193},
	issn = {0010-2180},
	doi = {10.1016/j.combustflame.2018.03.021},
	urldate = {2024-05-22},
	journal = {Combustion and Flame},
	author = {Xu, Rui and Wang, Kun and Banerjee, Sayak and Shao, Jiankun and Parise, Tom and Zhu, Yangye and Wang, Shengkai and Movaghar, Ashkan and Lee, Dong Joon and Zhao, Runhua and Han, Xu and Gao, Yang and Lu, Tianfeng and Brezinsky, Kenneth and Egolfopoulos, Fokion N. and Davidson, David F. and Hanson, Ronald K. and Bowman, Craig T. and Wang, Hai},
	month = jul,
	year = {2018},
	pages = {520--537},
}

@article{nguyen_spontaneous_2019,
	title = {Spontaneous and {Triggered} {Longitudinal} {Combustion} {Instability} in a {Single}-{Injector} {Liquid}-{Rocket} {Combustor}},
	volume = {57},
	issn = {0001-1452},
	doi = {10.2514/1.J057743},
	number = {12},
	urldate = {2025-05-20},
	journal = {AIAA Journal},
	author = {Nguyen, Tuan M. and Sirignano, William A.},
	month = dec,
	year = {2019},
	pages = {5351--5364},
}

@article{shadram_neural_2021,
	title = {Neural {Network} {Flame} {Closure} for a {Turbulent} {Combustor} with {Unsteady} {Pressure}},
	volume = {59},
	issn = {0001-1452},
	doi = {10.2514/1.J059721},
	number = {2},
	urldate = {2025-05-20},
	journal = {AIAA Journal},
	author = {Shadram, Zeinab and Nguyen, Tuan M. and Sideris, Athanasios and Sirignano, William A.},
	month = feb,
	year = {2021},
	pages = {621--635},
}

@article{chien_predictions_1982,
	title = {Predictions of {Channel} and {Boundary}-{Layer} {Flows} with a {Low}-{Reynolds}-{Number} {Turbulence} {Model}},
	volume = {20},
	issn = {0001-1452},
	doi = {10.2514/3.51043},
	number = {1},
	urldate = {2025-12-08},
	journal = {AIAA Journal},
	author = {Chien, Kuei-Yuan},
	month = jan,
	year = {1982},
	pages = {33--38},
}

@article{arts_aero-thermal_1992,
	title = {Aero-{Thermal} {Performance} of a {Two}-{Dimensional} {Highly} {Loaded} {Transonic} {Turbine} {Nozzle} {Guide} {Vane}: {A} {Test} {Case} for {Inviscid} and {Viscous} {Flow} {Computations}},
	volume = {114},
	issn = {0889-504X},
	shorttitle = {Aero-{Thermal} {Performance} of a {Two}-{Dimensional} {Highly} {Loaded} {Transonic} {Turbine} {Nozzle} {Guide} {Vane}},
	doi = {10.1115/1.2927978},
	number = {1},
	urldate = {2025-12-11},
	journal = {Journal of Turbomachinery},
	author = {Arts, T. and Lambert de Rouvroit, M.},
	month = jan,
	year = {1992},
	pages = {147--154},
}

@incollection{liu_computational_2025,
	address = {Reston, VA},
	title = {Computational fluid dynamics for turbomachinery},
	booktitle = {Gas {Turbine} {Compressors} and {Fans}: {Fundamentals}, {Design}, and {Analysis}},
	publisher = {AIAA Progress in Astronautics and Aeronautics},
	author = {Liu, Feng and Zhu, Yalu},
	editor = {Wellborn, S. R. and Shih, T. I.-P. and Yang, V.},
	year = {2025},
	doi = {10.2514/5.9781624107221.0193.0366},
	note = {Section: 5},
	pages = {193--366},
}

@inproceedings{zhu_numerical_2017,
	title = {Numerical {Investigation} of {Stator} {Clocking} {Effects} on the {Downstream} {Stator} in a 1.5-{Stage} {Axial} {Turbine}},
	doi = {10.1115/GT2017-63273},
	booktitle = {{ASME} {Turbo} {Expo} 2017: {Turbomachinery} {Technical} {Conference} and {Exposition}, {GT2017}-63273},
	author = {Zhu, Yalu and Luo, Jiaqi and Liu, Feng},
	month = jun,
	year = {2017},
}

@incollection{jameson_numerical_1981,
	series = {Fluid {Dynamics} and {Co}-located {Conferences}},
	title = {Numerical solution of the {Euler} equations by finite volume methods using {Runge} {Kutta} time stepping schemes},
	urldate = {2025-12-08},
	booktitle = {14th {Fluid} and {Plasma} {Dynamics} {Conference}},
	publisher = {American Institute of Aeronautics and Astronautics},
	author = {Jameson, A. and Schmidt, Wolfgang and Turkel, Eli},
	month = jun,
	year = {1981},
}

@article{zhu_flow_2018,
	title = {Flow {Computations} of {Multi}-{Stages} by {URANS} and {Flux} {Balanced} {Mixing} {Models}},
	volume = {61},
	doi = {10.1007/s11431-017-9262-9},
	journal = {Science China Technological Sciences},
	author = {Zhu, Yalu and Luo, Jiaqi and Liu, Feng},
	year = {2018},
	pages = {1081--1091},
}

@article{zhu_influence_2018,
	title = {Influence of {Blade} {Lean} {Together} with {Blade} {Clocking} on the {Overall} {Aerodynamic} {Performance} of a {Multi}-{Stage} {Turbine}},
	volume = {80},
	doi = {10.1016/j.ast.2018.07.016},
	journal = {Aerospace Science and Technology},
	author = {Zhu, Yalu and Luo, Jiaqi and Liu, Feng},
	year = {2018},
	pages = {329--336},
}

@article{sirignano_three-dimensional_2022,
	title = {Three-dimensional, rotational flamelet closure model with two-way coupling},
	volume = {945},
	issn = {0022-1120, 1469-7645},
	doi = {10.1017/jfm.2022.562},
	language = {en},
	urldate = {2025-12-03},
	journal = {Journal of Fluid Mechanics},
	author = {Sirignano, William A.},
	month = aug,
	year = {2022},
	pages = {A21},
}

@article{sirignano_inward_2022,
	title = {Inward swirling flamelet model},
	volume = {26},
	issn = {1364-7830},
	doi = {10.1080/13647830.2022.2103452},
	number = {6},
	urldate = {2025-12-03},
	journal = {Combustion Theory and Modelling},
	author = {Sirignano, William A.},
	month = sep,
	year = {2022},
	pages = {1014--1040},
}

@article{hellwig_vortex_2025,
	title = {Vortex stretching of non-premixed, diluted hydrogen/oxygen flamelets},
	volume = {273},
	issn = {0010-2180},
	doi = {10.1016/j.combustflame.2024.113900},
	urldate = {2025-12-03},
	journal = {Combustion and Flame},
	author = {Hellwig, Wes and Shi, Xian and Sirignano, William A.},
	month = mar,
	year = {2025},
	pages = {113900},
}

@article{hellwig_three-dimensional_2025,
	title = {Three-dimensional vorticity effects on extinction behaviour of laminar flamelets},
	volume = {29},
	issn = {1364-7830},
	doi = {10.1080/13647830.2024.2438778},
	number = {1},
	urldate = {2025-12-03},
	journal = {Combustion Theory and Modelling},
	author = {Hellwig, Wes and Shi, Xian and Sirignano, William A.},
	month = jan,
	year = {2025},
	pages = {62--92},
}

@article{tao_critical_2018,
	series = {Special {Commemorative} {Issue}: {Professor} {Chung} {King} ({Ed}) {Law} 70th {Birthday}},
	title = {Critical kinetic uncertainties in modeling hydrogen/carbon monoxide, methane, methanol, formaldehyde, and ethylene combustion},
	volume = {195},
	issn = {0010-2180},
	doi = {10.1016/j.combustflame.2018.02.006},
	urldate = {2025-05-21},
	journal = {Combustion and Flame},
	author = {Tao, Yujie and Smith, Gregory P. and Wang, Hai},
	month = sep,
	year = {2018},
	pages = {18--29},
}

@misc{smith_foundational_2016,
	title = {Foundational {Fuel} {Chemistry} {Model} {Version} 1.0 ({FFCM}-1)},
	url = {http://nanoenergy.stanford.edu/ffcm1},
	author = {Smith, G. P. and Tao, Y. and Wang, H.},
	year = {2016},
}

@article{westbrook_chemical_1984,
	title = {Chemical kinetic modeling of hydrocarbon combustion},
	volume = {10},
	issn = {0360-1285},
	doi = {10.1016/0360-1285(84)90118-7},
	number = {1},
	urldate = {2025-05-21},
	journal = {Progress in Energy and Combustion Science},
	author = {Westbrook, Charles K. and Dryer, Frederick L.},
	month = jan,
	year = {1984},
	pages = {1--57},
}

@article{saghafian_efficient_2015,
	title = {An efficient flamelet-based combustion model for compressible flows},
	volume = {162},
	issn = {0010-2180},
	doi = {10.1016/j.combustflame.2014.08.007},
	number = {3},
	urldate = {2025-05-20},
	journal = {Combustion and Flame},
	author = {Saghafian, Amirreza and Terrapon, Vincent E. and Pitsch, Heinz},
	month = mar,
	year = {2015},
	pages = {652--667},
}

@article{shadram_physics-aware_2022,
	title = {Physics-aware neural network flame closure for combustion instability modeling in a single-injector engine},
	volume = {240},
	issn = {0010-2180},
	doi = {10.1016/j.combustflame.2021.111973},
	urldate = {2025-05-20},
	journal = {Combustion and Flame},
	author = {Shadram, Zeinab and Nguyen, Tuan M. and Sideris, Athanasios and Sirignano, William A.},
	month = jun,
	year = {2022},
	pages = {111973},
}

@article{nguyen_impacts_2018,
	series = {Special {Commemorative} {Issue}: {Professor} {Chung} {King} ({Ed}) {Law} 70th {Birthday}},
	title = {The impacts of three flamelet burning regimes in nonlinear combustion dynamics},
	volume = {195},
	issn = {0010-2180},
	doi = {10.1016/j.combustflame.2018.03.031},
	urldate = {2025-05-20},
	journal = {Combustion and Flame},
	author = {Nguyen, Tuan M. and Sirignano, William A.},
	month = sep,
	year = {2018},
	pages = {170--182},
}

@article{zhan_combustion_2024,
	title = {Combustion dynamics of ten-injector rocket engine using flamelet progress variable},
	volume = {267},
	issn = {0010-2180},
	doi = {10.1016/j.combustflame.2024.113538},
	urldate = {2025-05-20},
	journal = {Combustion and Flame},
	author = {Zhan, Lei and Nguyen, Tuan M. and Xiong, Juntao and Liu, Feng and Sirignano, William A.},
	month = sep,
	year = {2024},
	pages = {113538},
}

@article{pecnik_reynolds-averaged_2012,
	title = {Reynolds-{Averaged} {Navier}-{Stokes} {Simulations} of the {HyShot} {II} {Scramjet}},
	volume = {50},
	doi = {10.2514/1.J051473},
	number = {8},
	journal = {Associate Fellow AIAA. AIAA JOURNAL},
	author = {Pečnik, René and Terrapon, Vincent E and Ham, Frank and Iaccarino, Gianluca and Pitsch, Heinz and Aiaa, Member},
	year = {2012},
}

@inproceedings{mehring_ignition_2001,
	address = {Reston, Virigina},
	title = {Ignition and flame studies for a turbulent accelerating transonic mixing layer},
	doi = {10.2514/6.2001-190},
	publisher = {American Institute of Aeronautics and Astronautics},
	author = {Mehring, Carsten and Liu, Feng and Sirignano, William},
	month = jan,
	year = {2001},
}

@inproceedings{cai_ignition_2001,
	address = {Reston, Virigina},
	title = {Ignition and flame studies for turbulent transonic mixing in a curved duct flow},
	doi = {10.2514/6.2001-189},
	publisher = {American Institute of Aeronautics and Astronautics},
	author = {Cai, Jinsheng and Icoz, Olgu and Liu, Feng and Sirignano, W.},
	month = jan,
	year = {2001},
}

@article{cheng_reacting_2009,
	title = {Reacting {Mixing}-{Layer} {Computations} in a {Simulated} {Turbine}-{Stator} {Passage}},
	volume = {25},
	doi = {10.2514/1.37739},
	number = {2},
	journal = {Journal of Propulsion and Power},
	author = {Cheng, Felix and Liu, Feng and Sirignano, William A.},
	month = mar,
	year = {2009},
	pages = {322--334},
}

@article{fang_ignition_2001,
	title = {Ignition and {Flame} {Studies} for an {Accelerating} {Transonic} {Mixing} {Layer}},
	volume = {17},
	doi = {10.2514/2.5844},
	number = {5},
	journal = {Journal of Propulsion and Power},
	author = {Fang, X. and Liu, F. and Sirignano, W. A.},
	month = sep,
	year = {2001},
	pages = {1058--1066},
}

@inproceedings{cai_combustion_2001,
	address = {Seattle},
	title = {Combustion in a {Transonic} {Flow} with {Large} {Axial} and {Transverse} {Pressure} {Gradients}},
	author = {Cai, J and Icoz, O. and Liu, F. and Sirignano, W. A.},
	year = {2001},
	pages = {1--9},
}

@article{cheng_nonpremixed_2007,
	title = {Nonpremixed {Combustion} in an {Accelerating} {Transonic} {Flow} {Undergoing} {Transition}},
	volume = {45},
	doi = {10.2514/1.31146},
	number = {12},
	journal = {AIAA Journal},
	author = {Cheng, Felix and Liu, Feng and Sirignano, William A.},
	month = dec,
	year = {2007},
	pages = {2935--2946},
}

@article{cheng_nonpremixed_2008,
	title = {Nonpremixed {Combustion} in an {Accelerating} {Turning} {Transonic} {Flow} {Undergoing} {Transition}},
	volume = {46},
	doi = {10.2514/1.35209},
	number = {5},
	journal = {AIAA Journal},
	author = {Cheng, Felix and Liu, Feng and Sirignano, William A.},
	month = may,
	year = {2008},
	pages = {1204--1215},
}

@article{sirignano_turbine_2012,
	title = {Turbine {Burners}: {Performance} {Improvement} and {Challenge} of {Flameholding}},
	volume = {50},
	doi = {10.2514/1.J051562},
	number = {8},
	journal = {AIAA Journal},
	author = {Sirignano, W. A. and Dunn-Rankin, D. and Liu, F. and Colcord, B. and Puranam, S.},
	month = aug,
	year = {2012},
	pages = {1645--1669},
}

@article{liu_turbojet_2001,
	title = {Turbojet and {Turbofan} {Engine} {Performance} {Increases} {Through} {Turbine} {Burners}},
	volume = {17},
	doi = {10.2514/2.5797},
	number = {3},
	journal = {Journal of Propulsion and Power},
	author = {Liu, F. and Sirignano, W. A.},
	month = may,
	year = {2001},
	pages = {695--705},
}

@article{sirignano_performance_1999,
	title = {Performance {Increases} for {Gas}-{Turbine} {Engines} {Through} {Combustion} {Inside} the {Turbine}},
	volume = {15},
	doi = {10.2514/2.5398},
	number = {1},
	journal = {Journal of Propulsion and Power},
	author = {Sirignano, W. A. and Liu, F.},
	month = jan,
	year = {1999},
	pages = {111--118},
}

@article{peters_laminar_1984,
	title = {Laminar diffusion flamelet models in non-premixed turbulent combustion},
	volume = {10},
	doi = {10.1016/0360-1285(84)90114-X},
	number = {3},
	journal = {Progress in Energy and Combustion Science},
	author = {Peters, N.},
	year = {1984},
	pages = {319--339},
}

@article{peters_local_1983,
	title = {Local {Quenching} {Due} to {Flame} {Stretch} and {Non}-{Premixed} {Turbulent} {Combustion}},
	volume = {30},
	doi = {10.1080/00102208308923608},
	number = {1-6},
	journal = {Combustion Science and Technology},
	author = {Peters, N.},
	month = jan,
	year = {1983},
	pages = {1--17},
}

\end{document}